\documentclass[iop]{emulateapj}
\usepackage{graphicx}
\usepackage{color}
\usepackage[normalem]{ulem}
\usepackage{wasysym}
\usepackage[bookmarks=true,
            bookmarkstype=toc,
            bookmarksopen=true,
            bookmarksopenlevel=10,
            pdfstartview = FitV,
            pdfpagelayout = SinglePage,
            hyperfigures=true,
            colorlinks = true, 
            linkcolor=blue,
            urlcolor = blue,
            citecolor = blue,
            pdfborder = {0 0 0}]{hyperref}

\shorttitle{X-ray Photo-polarimetry of HD~189733\upshape{b}}
\shortauthors{Marin \& Grosso}

\begin{document}
\title{Computation of the transmitted and polarized scattered fluxes\\
by the exoplanet HD~189733\MakeLowercase{b} in X-rays}

\author{Fr\'ed\'eric Marin$^{1,2}$ and Nicolas Grosso$^2$}
\affil{$^{1}$Astronomical Institute of the Academy of Sciences, Bo{\v c}n\'{\i} II 1401, CZ-14100 Prague, Czech Republic\\
$^{2}$Universit\'e de Strasbourg, CNRS, Observatoire astronomique de Strasbourg, UMR 7550, F-67000 Strasbourg, France} 
\email{frederic.marin@astro.unistra.fr}

\begin{abstract}
{Thousands of exoplanets have been detected, but only one exoplanetary transit was potentially observed in X-rays 
from HD~189733A. What makes the detection of exoplanets so difficult in this band? To answer this question, we run 
Monte-Carlo radiative transfer simulations to estimate the amount of X-ray flux reprocessed by HD~189733b
Despite its extended evaporating-atmosphere, we find that the X-ray absorption radius of HD~189733b at 0.7~keV, the 
mean energy of the photons detected in the 0.25--2 keV energy band by XMM-Newton, is $\sim$1.01 times the planetary 
radius for an atmosphere of atomic Hydrogen and Helium (including ions), and produces a maximum depth of $\sim$2.1\% 
at $\sim$$\pm46$~min from the center of the planetary transit on the geometrically thick and optically thin 
corona. We compute numerically in the 0.25--2~keV energy band that this maximum depth is only of 
$\sim$1.6\% at $\sim$$\pm47$~min from the transit center, and little sensitive to the metal abundance 
assuming that adding metals in the atmosphere would not dramatically change the density-temperature profile.
Regarding a direct detection of HD~189733b in X-rays, we find that the amount of flux reprocessed by the exoplanetary 
atmosphere varies with the orbital phase, spanning between 3--5 orders of magnitude fainter than the flux of the primary 
star. Additionally, the degree of linear polarization emerging from HD 189733b is $<$0.003\%, with maximums detected 
near planetary greatest elongations. This implies that both the modulation of the X-ray flux with the orbital phase and 
the scattered-induced continuum polarization cannot be observed with current X-ray facilities.}
\end{abstract}

\section{Introduction}
\label{Intro}

Hitherto, the main exoplanet detection technique relies on spectroscopic and photometric transits\footnote{In December 2016, 2695 
planets out of a total of 3547 have been discovered by transit techniques, see http://exoplanet.eu/}. If a planet passes in front of a star, 
the star will be partially eclipsed and its flux will be dimmed by the successive coverage and uncoverage of the stellar surface by the transiting 
planet, resulting in detectable dips in the stellar light curve. In the optical and infrared bands, where most of the transit detection 
occurred, those transit depths are of the order of 1--3\%, e.g., 1.6\% for HD~209458b \citep{Charbonneau2000,Brown2001}, and 2.45\% for 
HD~189733b \citep{Lecavelier2008,Bouchy2005}. If the photospheric disk has a uniform emission (i.e., there is no limb-darkening), these depths
are equal to the square of the ratio of the exoplanet radius ($R_\mathrm{p}$) to the stellar radius ($R_\star$). From multi-wavelength 
light curves, it becomes possible to characterize the atmosphere and exosphere of giant exoplanets \citep[e.g.,][]{Redfield2008},
but additional measurements are needed to estimate the mass of the planet, the orbital parameters, and the inclination angle. These results
can be achieved thanks to radial velocity measurements, orbital brightness modulations, gravitational microlensing, direct imaging, astrometry, 
and polarimetry. The later is probably one of the most delicate methods since the anticipated white light polarization degree (about 10$^{-5}$, 
\citealt{Carciofi2005,Kostogryz2014,Kopparla2016}) resulting from scattering of stellar photons onto the atmospheric layers of Hot Jupiters is 
expected to be just about detectable given the current limits of current optical, broad-band, stellar polarimeters \citep{Schmid2005,Hough2006,Wiktorowicz2009}. 
\citet{Berdyugina2008,Berdyugina2011} reported to have successfully achieved a $B$-band polarimetric detection of HD~189733b and found a peak
polarization of 2$\times$10$^{-4}$ (i.e. 0.02~\%) and a cosine-shaped distribution of polarization over the orbital period. From those results, 
\citet{Berdyugina2008} inferred additional constraints on the size of the scattering atmosphere of HD~189733b, and on the orientation of the 
system with respect to the Earth-primary line of sight. Yet there is no reliable measurement of the true scattered light polarization 
from HD~189733b in the $B$-band, as \citet{Wiktorowicz2015} and \citet{Bott2016} reported an absence of large amplitude polarization 
variations. Even if their observations are consistent with a plausible polarization amplitude from the planet, the low significance of the 
polarimetric observations cannot be used as a claim for detection. This highlights the difficulty of \textit{polarimetric} measurements 
of exoplanets in the optical band. However, \textit{spectroscopic} detection of exoplanets has been quite successful at all wavebands, 
except at X-rays energies.

It is unfortunate since X-ray photometry, spectroscopy, and polarimetry could help to constrain the composition and morphology of the 
exoplanet's atmosphere. \citet{Poppenhaeger2013} reported for HD~189733 the detection in the 0.25--2~keV energy band of a transit depth
estimate ranging from 4.0$\pm$2.7\% to 7.3$\pm$2.5\% (see in their Table~6 the limb-brightened model), which might be larger than the 
transit depth in optical of 2.391$\pm$0.001\%, computed for a uniformly emitting disk using the planetary radius 
of \citet{Torres2008}. These values of transit depths in X-rays were obtained by fitting the X-ray broad-band light curve 
using the analytic chromospheric transit of \citet{Schlawin2010} that assumes a geometrically thin emission. However, the 
corona of the Sun (a normal star) is by definition located above the chromosphere and the solar coronal plasma is extended above 
the photosphere \citep[see for a review on the solar corona, e.g.,][]{Aschwanden2004}, therefore, a zero thickness is not appropriate 
to describe the geometry of the corona of HD~189733A, a moderately active star. Moreover, the normalized phase-folded X-ray light 
curve of HD~189733A is affected for a phase binning of 0.005 (corresponding to 958~s) by large statistical noise ($\sim$5\%; 
Poisson noise from added-up raw count rate out transit in Table~5 of \citealt{Poppenhaeger2013}) and astrophysical (white noise) scatter 
($\sim$4\%; see bottom right panel of Fig.~12 in the Appendix of \citealt{Poppenhaeger2013}) that hinders any fitting.
Therefore, more sensitive X-ray observations of the transit of HD~189733b are mandatory to obtain an accurate measurement 
of the shape of the planetary transit of the corona of HD~189733A and to better constrain its maximum depth.

In any case, the first indirect detection of an extra-solar planet in the 0.1 -- 10~keV energy 
range opens new questions: What is the expected X-ray absorption radius and scattered flux from giant exoplanets? Is a \textit{direct} 
detection of giant exoplanets feasible in this energy band? Is, similarly to the optical band, a polarimetric detection within the reach 
of future instruments? To solve those issues, it is necessary to estimate the host-star flux reprocessed by the exoplanet gaseous surface 
first, and then to compute the resulting scattering-induced polarization. It is a particularly relevant issue since we have known for 
decades that stellar X-ray polarimetry is a very powerful diagnostic tool \citep[e.g.,][]{Manzo1993}. 

In this regards, HD~189733b is one of the best laboratories for giant exoplanet studies. HD~189733 is a binary system, situated at a 
distance of about 19.45~pc from the Sun \citep{VanLeeuwen2007}. The primary component HD~189733A is a main-sequence star of stellar type 
K1.5~V, mass 0.846~$M_\odot$ \citep{Kok2013}, radius 0.805~$R_\odot$ \citep{Belle2009,Boyajian2015} and rotational period 11.953 days 
\citep{Henry2008}. HD~189733A emits about 10$^{28~}$erg.s$^{-1}$ in the 0.25 -- 2~keV band, corresponding to $10^{-5}$ times 
its bolometric luminosity, which is about 10 times larger than the ratio of X-ray to bolometric luminosity observed from the Sun at 
its maximum \citep{Poppenhaeger2013}. The secondary component of spectral type M4~V is located at an angular separation of 11$\farcs$4 
($\sim$ 216~au) from HD~189733A with an orbital period is 3200~years \citep{Bakos2006} and it is approximatively two 
orders of magnitude fainter in X-rays than HD~189733A \citep{Poppenhaeger2013}. The exoplanet HD~189733b with mass 1.162~$M_\mathrm{J}$
\citep{Kok2013} and radius 1.26~$R_\mathrm{J}$, was discovered by transit observations in optical \citep{Bouchy2005}. Orbiting around 
HD~189733A in 2.219 days \citep{Triaud2009}, this Hot Jupiter is situated at only 0.031~au from its host star. Due to its proximity 
from Earth and its favorable observing geometrical conditions --- its orbital plane is parallel within about 4$^{\circ}$ of our line of 
sight \citep{Triaud2009} and its orbit is nearly circular \citep{Berdyugina2008} --- HD~189733b stands as the perfect object for 
observations and numerical modeling.

Hence, in this paper, we aim to investigate the HD~189733A planetary system in terms of photometry and continuum polarization. Focusing 
our study on soft X-rays, where the stellar coronal emission peaks, we build in Sect.~\ref{Model} a template X-ray spectrum for the corona 
of HD~189733A and a model for the atmosphere of HD~189733b according to the latest observational constraints. We present the Monte-Carlo code 
that we used to perform our calculations of radiative transfer and report the very first estimations of the reprocessed flux emerging 
from the outer atmospheric layers of HD~189733b, along with its resulting X-ray polarization, in Sect.~\ref{Results}. In Sect.~\ref{Discussion}, 
we discuss our results in the context of present and future X-ray facilities, and conclude our paper in Sect.~\ref{Conclusion}.

\begin{figure}[!t]
  \centering
  \includegraphics[trim = 35mm 0mm 5mm 5mm, clip, width=9.7cm]{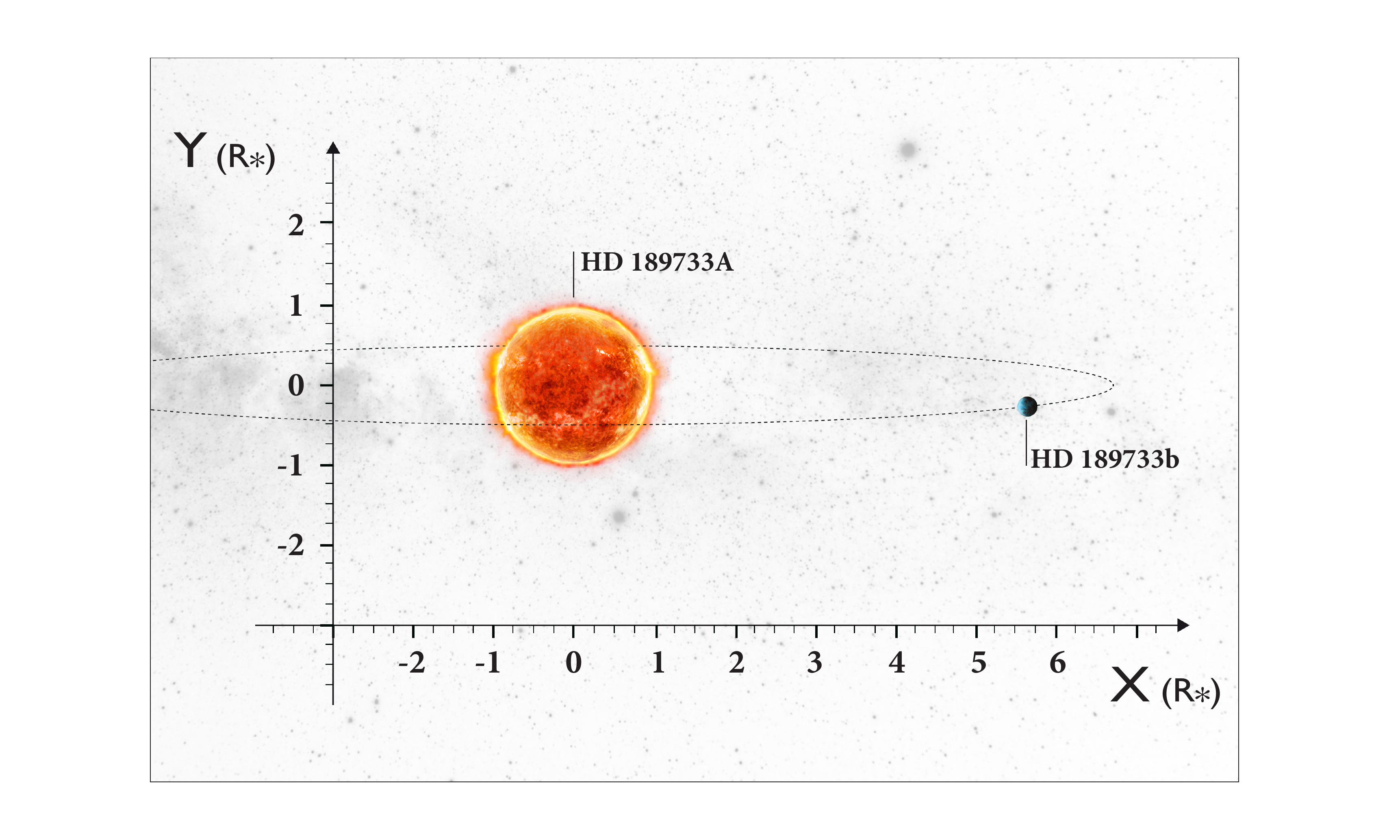} 
  \caption{Artist representation of the exoplanet HD~189733b. HD~189733b
	   orbits in nearly circular motion around its host-star HD~189733A;
	   its orbital plane is parallel within about 4$^{\circ}$ of the 
	   Earth-primary line of sight. Note the limb brightening effect 
	   emerging from the optically thin, stellar corona. }
  \label{Fig:Sketch} 
\end{figure}

\section[Modeling the HD~189733A corona and the HD~189733b atmosphere]{Modeling the HD~189733A corona and the HD~189733\MakeLowercase{b} atmosphere}
\label{Model}

Probing the gases and plasmas that form exoplanet atmospheres is a challenging task, as extra-solar bodies are usually very faint. 
Yet, constraints on the atmospheric composition and temperature of the external layers of hot Jupiter and hot Neptune-like planets can be 
derived from primary and secondary eclipses for transiting exoplanets \citep{Seager2000}. Numerical models are thus very useful to evaluate 
the detection capabilities of our current and forthcoming generation of telescopes, together with fitting observed spectra with atmospheric 
models. It is in this scope that we now present a planetary system model for HD~189733 derived from observations. A sketch of our model is 
presented in Fig.~\ref{Fig:Sketch}. We list in Tab.~\ref{Tab:Model} the physical properties of HD~189733A and HD~189733b that we use in this 
work. Note that our results are likely to vary if a new set of input parameters is derived from future observations. 

\subsection{Template X-ray spectrum of the HD~189733A quiescent corona}
\label{Model:Corona}

\begin{deluxetable}{@{}lll@{}}
  \tabletypesize{\footnotesize}
  \tablecolumns{3}
  \centering
  \tablewidth{\columnwidth}
  \tablecaption{Physical properties of HD~189733A and HD~189733b}
  \tablehead{\colhead{Property} & \colhead{Value} & \colhead{Reference}}
  \startdata 
	$R_\star/R_\odot$				& $0.756\pm0.018$ 						& \citet{Torres2008}\tablenotemark{a}\\
	$M_\star/M_\odot$				& $0.806\pm0.048$						& \citet{Torres2008}\tablenotemark{b}\\
	$R_\mathrm{p}/R_\star$				& $0.15463\pm0.00022$ 						& \citet{Torres2008}\tablenotemark{b}\\
	$R_\mathrm{p}/R_\mathrm{J}$\tablenotemark{d}	& $1.138 \pm 0.027$						& \citet{Torres2008}\tablenotemark{b}\\
	$M_\mathrm{p}/M_\mathrm{J}$\tablenotemark{e}			& $1.144^{+0.057}_{-0.056}$					& \citet{Torres2008}\tablenotemark{b}\\
	$a$/au						& $0.03099^{+0.00060}_{-0.00063}$	 			& \citet{Torres2008}\tablenotemark{b}\\
	$R_\mathrm{Roche}/R_\mathrm{p}$			& 4.35								& \citet{Salz2016}\\
	$P/\mathrm{days}$				& $2.21857312^{+0.00000036}_{-0.00000076}$			& \cite{Triaud2009}\\
	$d/\mathrm{pc}$					& 19								& \citet{Bouchy2005}\tablenotemark{b,c}\\
	$i_\mathrm{orbit}/\mathrm{deg}$			& $85.58\pm0.006$ 						& \citet{Torres2008}\\
	$b$						& $0.680\pm0.005$						& \citet{Torres2008}
	\enddata
   \tablenotetext{a}{\cite{Boyajian2015} obtained $0.805\pm0.016$ from interferometric measurement 
   using the distance of 19.45~pc \citep{VanLeeuwen2007}.}
   \tablenotetext{b}{We use this reference to be consistent with the physical properties used by 
   \citet{Salz2015} to compute their atmospheric model; see the header of the file {\tt hd189733.dat} 
   available at {\tt ftp://cdsarc.u-strasbg.fr/pub/cats/J/A\%2BA/586/A75/}\,.}
   \tablenotetext{c}{Works reporting X-ray observations used a distance of  
   19.3~pc \citep{Pillitteri2010,Pillitteri2011,Pillitteri2014,Poppenhaeger2013}.
   \cite{VanLeeuwen2007} reported a distance of $19.45\pm0.26$~pc from the new reduction 
   of the Hipparcos parallax.} 
   \tablenotetext{d}{$R_\mathrm{J}$ is the nominal jovian equatorial radius
   following IAU's recommandation (${\cal R}^\mathrm{N}_\mathrm{eJ}=7.1492\times10^7$~m;
   \citealt{Mamajek2015}).}   
   \tablenotetext{e}{The jovian mass is computed following 
   IAU's recommandation \citep{Mamajek2015}: 
   $M_\mathrm{J}=({\cal GM})^\mathrm{N}_\mathrm{J}/G$
   with (${\cal GM})^\mathrm{N}_\mathrm{J}=1.266\,865\,3\times10^{23}$~cm$^3$~s$^{-2}$ 
   the nominal jovian mass and $G=6.67384\times10^{-11}$~cm$^3$~kg$^{-1}$~s$^{-2}$ 
   the Newtonian constant.}
  \label{Tab:Model}
\end{deluxetable}

The moderately active corona of HD~189733A emits X-ray photons from an optically thin plasma in collisional ionization-equilibrium; 
the X-ray spectrum is a bremsstrahlung continuum emission plus line emission from metals. The X-ray spectra of the quiescent 
corona obtained with XMM-Newton and Chandra are well described by two-isothermal plasma 
\citep{Pillitteri2010,Pillitteri2011,Pillitteri2014,Poppenhaeger2013}. Each isothermal plasma is characterized by an electronic 
temperature and a corresponding emission measure, defined as $EM=\int n_\mathrm{e}^2 dV$ for a fully ionized plasma where the
electronic density, $n_\mathrm{e}$, is equal to the ion density. The photoelectric absorption above 0.2~keV along the line of sight 
is negligible since the distance of this star is low. 

We use the spectral fitting parameters of the quiescent corona obtained on April 17, 2007 with XMM-Newton by \cite{Pillitteri2011}
for a distance of $d=19.3$~pc: a cool plasma component with $kT_1=0.24\pm0.01$~keV (corresponding to $T_\mathrm{e,1}=2.8\pm0.1$~MK) 
and $EM_1=4.7_{-0.3}^{+0.4}\times10^{50}$~cm$^{-3}$, and warm plasma component with $kT_2$=0.71$_{-0.03}^{+0.04}$~keV (corresponding 
to $T_\mathrm{e,2}=8.2_{-0.3}^{+0.5}$~MK) and $EM_2=(2.8\pm0.3)\times10^{50}$~cm$^{-3}$.

\begin{figure}[!t]
  \centering
  \includegraphics[width=0.95\columnwidth]{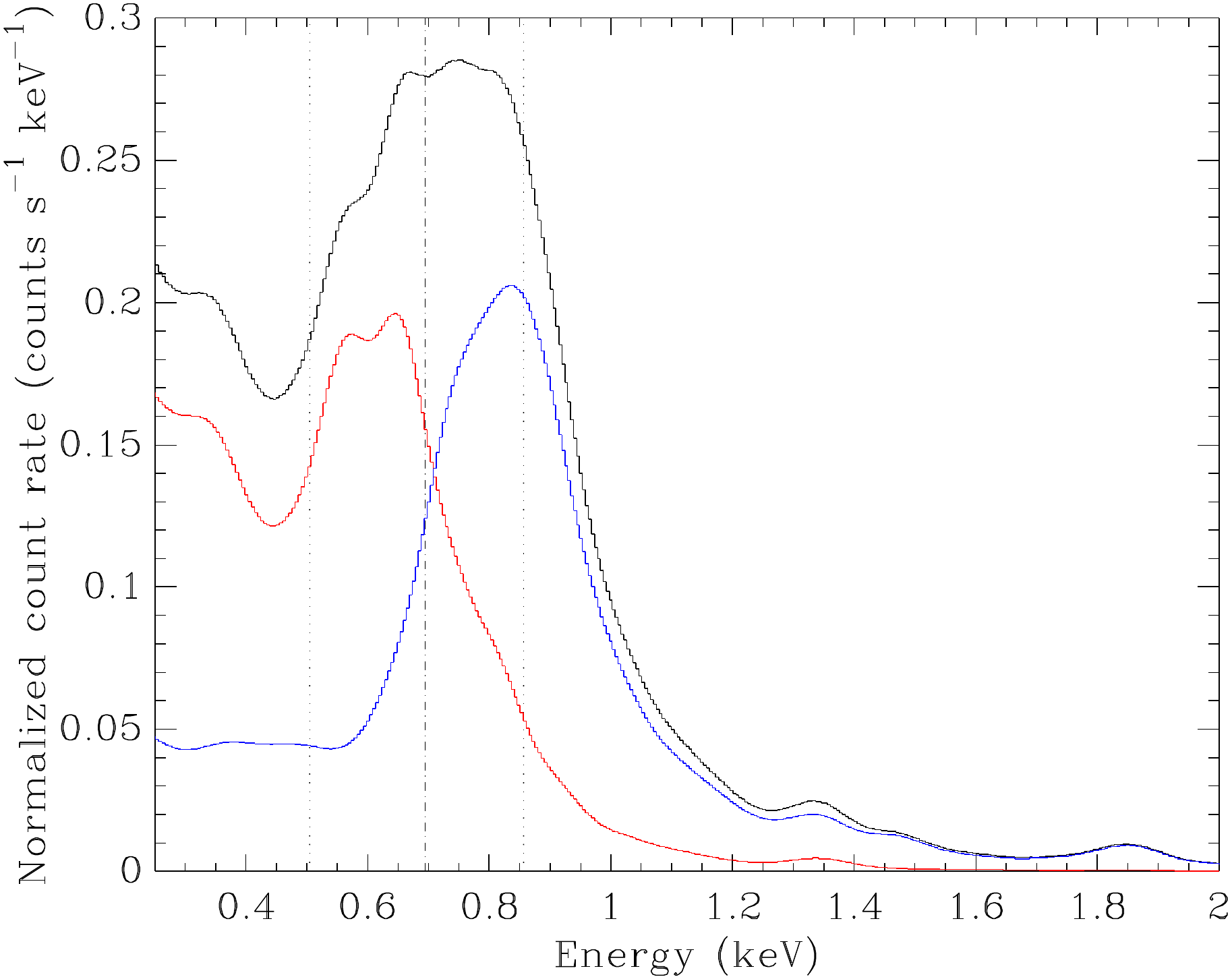} 
  \includegraphics[width=0.95\columnwidth]{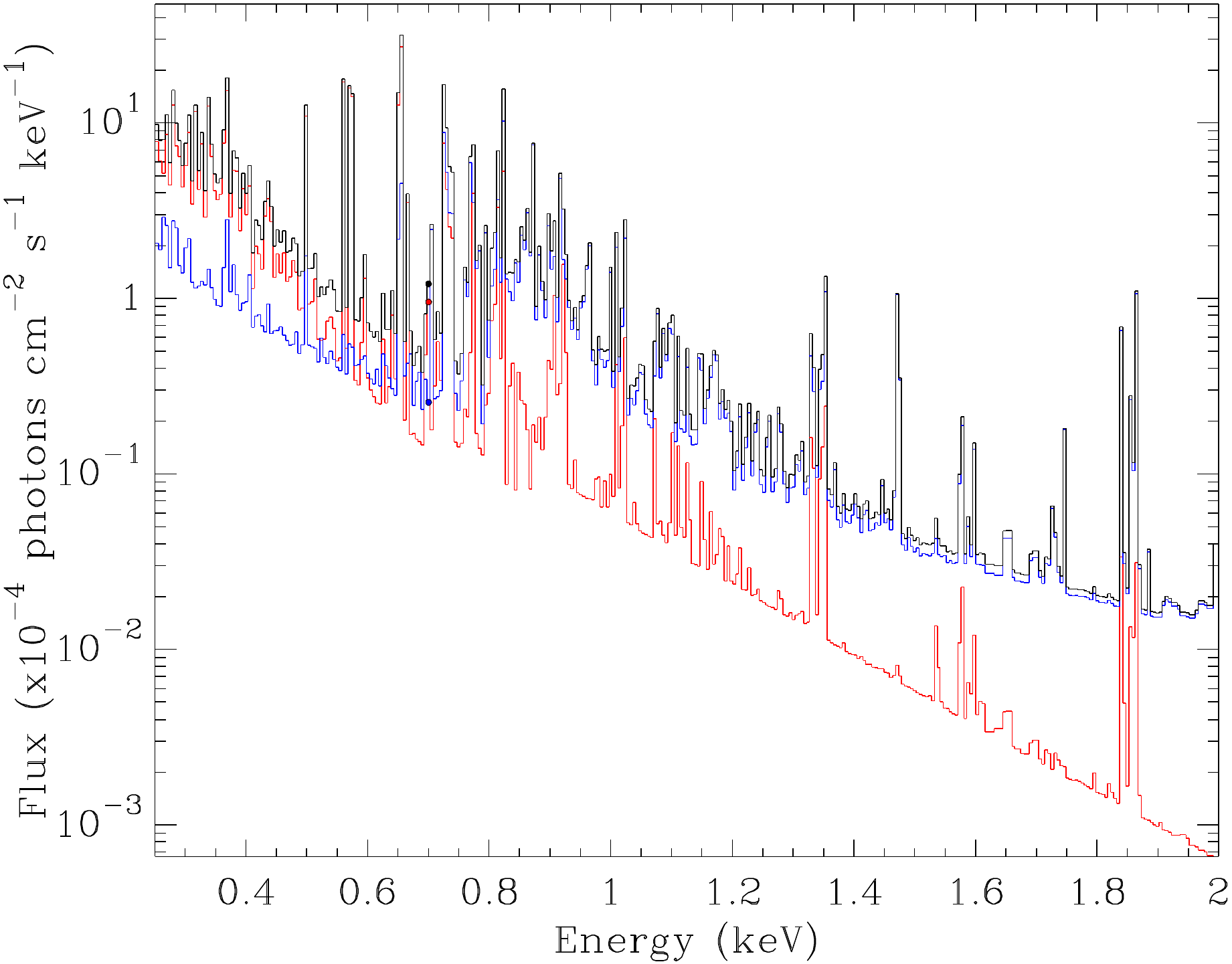} 
  \caption{XMM-Newton/pn template spectrum of HD~189733A 
	  used in this work. The energy range is 0.25--2~keV. 
	  Top panel: instrumental spectrum 
	  simulated from X-ray observations, grouped 
	  with a minimum of 25 counts. The dashed-dotted and 
	  dotted vertical lines are the energies of the median 
	  (50th-percentile), lower and upper quantiles (25th 
	  and 75th-percentiles) of the observed count rates, 
	  respectively. The red, blue, and black lines are the 
	  cool, warm, and total plasma components, respectively. 
	  Bottom panel: unfolded spectrum. The red, blue, 
	  and black dots are the cool, warm, and total plasma 
	  components used for our modeling of the X-ray corona, 
	  respectively.}
  \label{Fig:pn_spectrum} 
\end{figure}

We model this quiescent coronal emission with {\tt XSPEC} (version 12.9.0) using two {\tt vapec} models \citep{Smith2001}
with element abundances of \cite{Anders1989} and Fe=0.57, O=0.51, and Ne=0.3 \citep{Pillitteri2011}. We simulate an XMM-Newton 
observation of this coronal model with the European Photon Imaging Camera (EPIC) pn instrument \citep{Struder2001}
and the medium filter\footnote{We use the most recent canned response matrix for pn ({\tt epn\_e2\_ff20\_sdY9\_v14$.$0$.$rmf}) 
and the ancillary response file for a~40\arcsec-radius extraction region centered on this on-axis source, obtained 
from the XMM-Newton Serendipitous Source Catalogue 3XMM-DR4 \citep{Watson2009}.} \citep{Struder2001} which determines in the 
0.25--2.0~keV energy range an X-ray flux of $2.9\times 10^{-13}$~erg.cm$^{-2}$.s$^{-1}$ (corresponding to an X-ray luminosity 
of $1.3\times 10^{28}$~erg.s$^{-1}$ for $d$=19.3~pc) with a count rate of 0.188~pn~count~s$^{-1}$.

The count rate distribution versus energy, i.e., the simulated instrumental spectrum, is shown in the top panel of
Fig.~\ref{Fig:pn_spectrum}. Among the total counts detected in the 0.25--2.0~keV energy range, 25\% have energy lower than
0.51~keV (see lower dotted vertical line), 50\% have energy lower than 0.69~keV (see lower dashed-dotted vertical line),
75\% have energy lower than 0.86~keV (see upper dotted vertical line). The bottom panel of Fig.~\ref{Fig:pn_spectrum} is 
the so-called unfolded spectrum with an energy sampling\footnote{Our energy grid is defined by the nominal ranges of energy 
for the detector channels that are listed in the {\tt EBOUNDS} extension of the response matrix file.} of $\sim$5~eV.

We will use this template spectrum to compute numerically the depth of the transit observed in the 0.25--2~keV energy 
range with XMM-Newton (Sect.~\ref{Model:Transit}).

We will perform in Sect.~\ref{Model:Code} Monte-Carlo radiative transfer simulations at the mean energy of the photons detected 
in the 0.25-2~keV energy band, $E_\mathrm{mean}=0.70$~keV, where the X-ray emission is mainly bremsstrahlung continuum emission.
The coronal photon flux at this energy is $11.8\times10^{-5}$~photons~cm$^{-2}$~s$^{-1}$, with $F_\mathrm{ph,1}=9.2\times10^{-5}$ and 
$F_\mathrm{ph,2}=2.6\times10^{-5}$~photons~cm$^{-2}$~s$^{-1}$, the photon flux from the cool and warm plasma, respectively (bottom panel of 
Fig.~\ref{Fig:pn_spectrum}). We will assess the impact on the transmitted flux of X-ray observation in a broad energy-band in 
Sect.~\ref{Sect:broad_band}.

\subsection{Model of the HD~189733 quiescent corona}
\label{Sect:corona}

\begin{figure}[!t]
  \centering
  \includegraphics[width=\columnwidth]{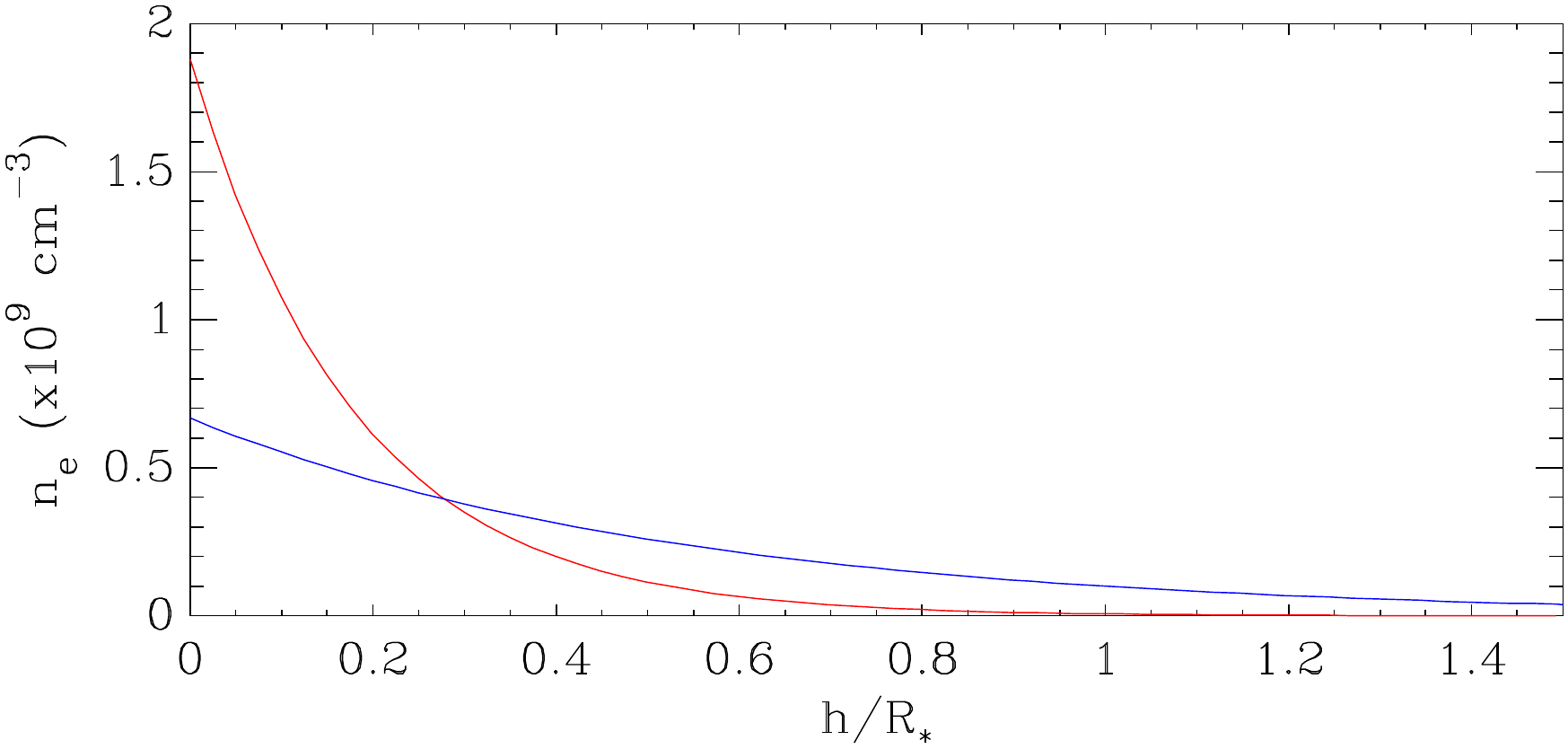} 
  \caption{Coronal electronic density versus the height above the corona. 
    The red and blue lines are for the cool and warm coronal plasma, 
    respectively.}
  \label{Fig:coronal_ne}
  \medskip
  \includegraphics[width=\columnwidth]{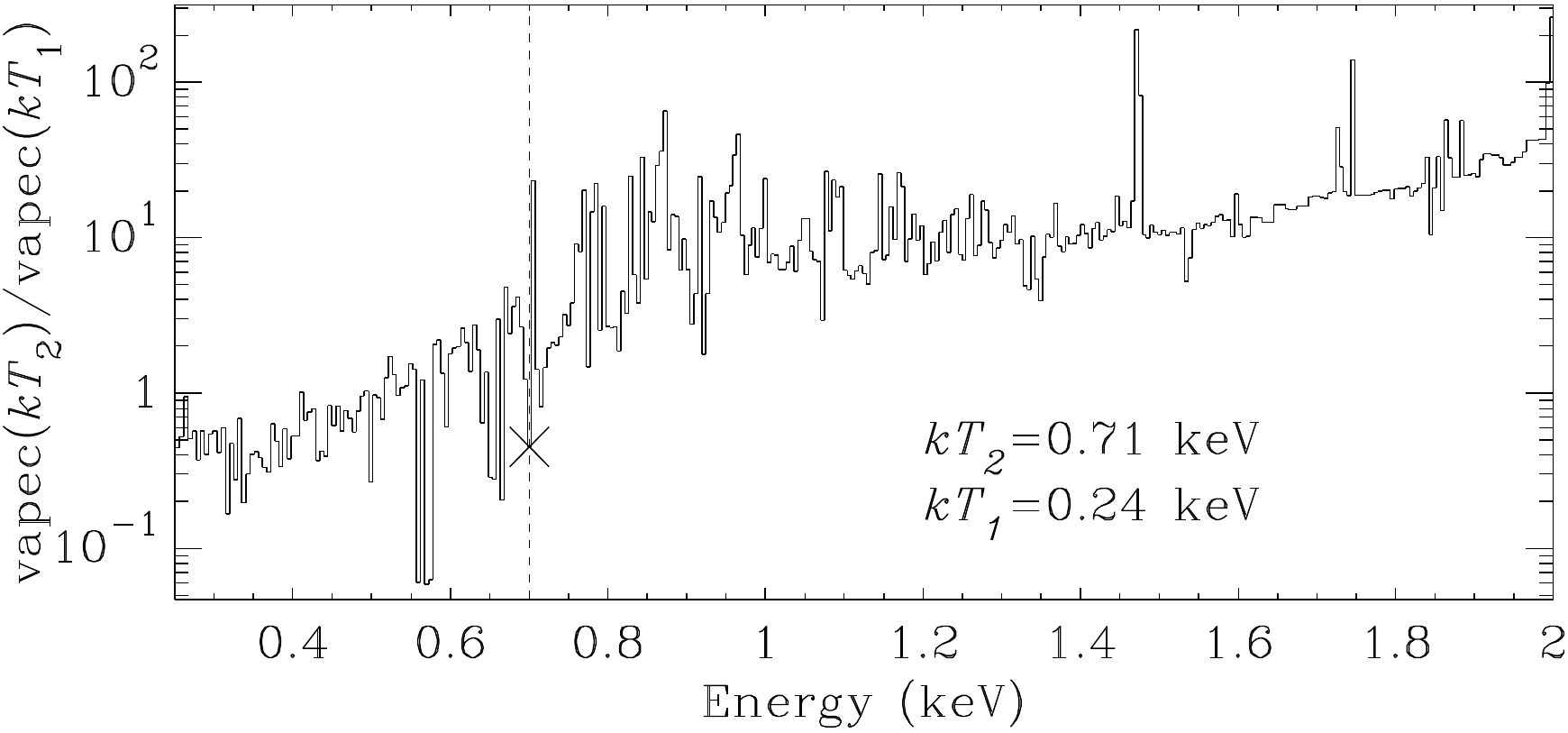}
  \caption{Ratio of warm-to-cool plasma emissivity (with identical 
    normalization) versus energy. The cross is the value corresponding to 
    $E_\mathrm{mean}=0.7$~keV (vertical dashed line).}
  \label{Fig:vapec_ratio} 
  \medskip
  \includegraphics[width=\columnwidth]{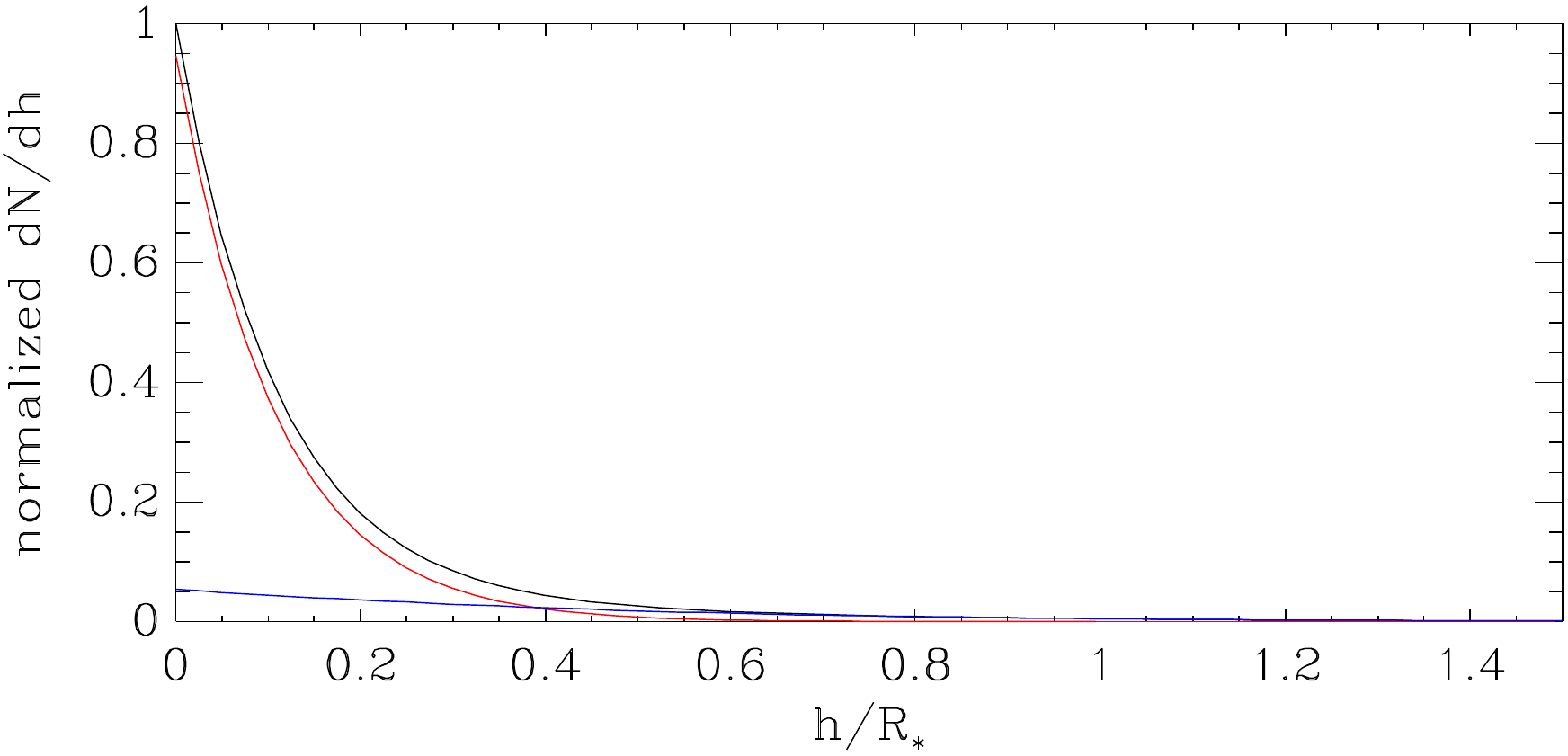} 
  \caption{Normalized photon emission at $E_\mathrm{mean}=0.7$~keV versus 
    the height above the corona. The red and blue are for the 
    cool and warm coronal plasma, respectively. The solid line is for 
    the cool+warm coronal plasma and is normalized at the peak.}
  \label{Fig:coronal_dN_dh} 
  \medskip
  \includegraphics[width=\columnwidth]{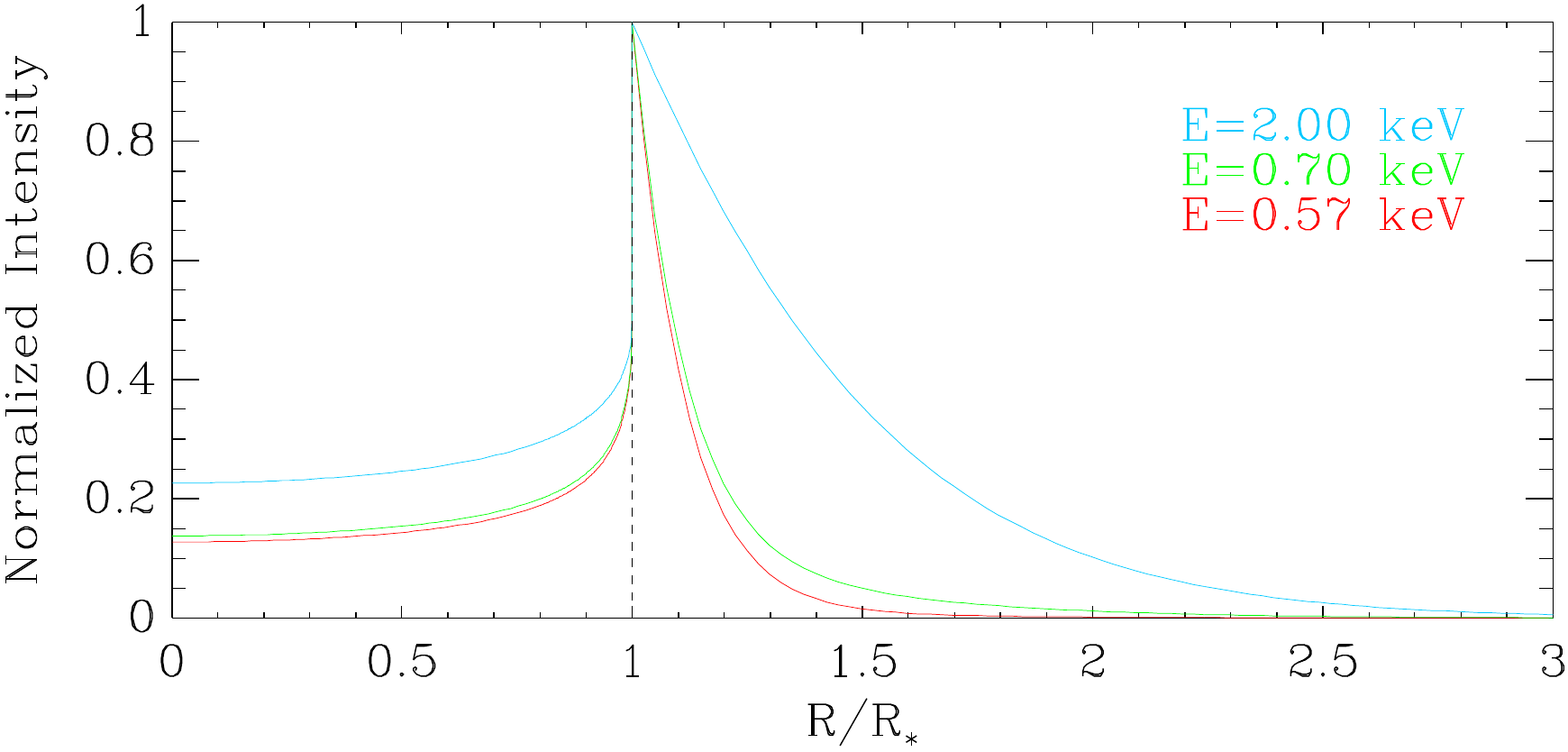} 
  \caption{Coronal emission profile integrated along the line of 
    sight versus the distance from the stellar center. The vertical dashed 
    line is the solar limb. All the profiles have been normalized at the solar
    limb (vertical dashed line). The red and blue lines are the most compact 
    and extended coronal emission profiles occuring at 0.57 and 2.00~keV, 
    respectively. The green line is the coronal emission profile at 0.7~keV 
    that we use in our simulation.}
  \label{Fig:coronal_emission_profile}
\end{figure}

The X-ray emitting plasma of the Sun corona is visible both in active regions where it is confined in magnetic loops
and in the most homogeneous parts of the quiet Sun corona where there is any detectable features of magnetic loops; both coronal 
regions extending well above the photosphere. For HD~189733A only the large-scale magnetic topology is accessible from Zeeman-Doppler 
imaging, \citet{Moutou2007} showed that ``{\it it is significantly more complex than that of the Sun ; it involves in particular a 
significant toroidal component and contributions from magnetic multipoles of order up to 5}". However, it is beyond the scope of 
this article to use this magnetic topology as a scaffold to build a model of the X-ray emitting plasma of the HD~189733A corona 
(see, e.g., \citealt{Jardine2006}). We focus on the quiescent and diffuse corona, which is more representative of the average plasma 
distribution on the stellar surface. Therefore, following our template spectrum, we model the quiescent corona as two isothermal-plasma 
in hydrostatic equilibrium (see Sect.\ 3.3 of \citealt{Aschwanden2004}). 

In the isothermal approximation, the pressure scale height is defined as:

\begin{equation}
  \lambda_\mathrm{p}(T_\mathrm{e})=\frac{2kT_\mathrm{e}}{\mu m_\mathrm{H}g_\star} \,,
  \label{eq:Lambdap}
\end{equation}

where $k$ is the Boltzmann's constant, $\mu\approx1.27$ is the mean particle weight, $m_\mathrm{H}$ is the mass of the hydrogen atom, 
and $g_\star=GM_\star/R_\star^2$ is the stellar gravitational field. 
For HD~189733, we obtain: 
$\lambda_\mathrm{p}(T_\mathrm{e})=3.4\times10^9\times(T_\mathrm{e}/\mathrm{MK})$~cm. Therefore, $\lambda_\mathrm{p}(T_\mathrm{e,1})=0.18 R_\star$ 
and $\lambda_\mathrm{p}(T_\mathrm{e,2})=0.53 R_\star$. Since the pressure is obtained by $p=2n_\mathrm{e}kT_\mathrm{e}$, the electronic
density scale height is identical to the pressure scale height in the isothermal approximation. Neglecting the variation of the gravitational 
field with the height above the photosphere, $h$, the electronic density, $n_\mathrm{e} $, follows an exponential profile:

\begin{equation}
  n_\mathrm{e}(h)=n_\mathrm{e}(0) 
  \exp{\!\left\{\frac{-h}{\lambda_\mathrm{p}(T_\mathrm{e})}\right\}} \,.
\end{equation}

We introduce, $f_\mathrm{obs}$, the fraction of the corona outer area that is not eclipsed by the stellar disk for an observer on 
Earth (Eq.~5 of \citealt{Schmitt1990}):

\begin{equation}
f_\mathrm{obs}(x)=0.5\times\frac{1+\sqrt{x(2+x)}}{1+x} \,,
\end{equation}

with $x=h/R_\star$. Obviously, only half of the coronal area is visible when $h=0$ ($f_\mathrm{obs}(0)=0.5$), and the coronal area is 
fully visible when $h=\infty$ ($f_\mathrm{obs}(\infty)=1.0$).

The observed emission measure must be computed on the coronal volume that is not eclipsed by the stellar disk:

\begin{equation}
EM_\mathrm{obs}=\int_0^\infty \frac{dEM(h)}{dh}
f_\mathrm{obs}\!\left(\frac{h}{R_\star}\right)\,dh \,,
\end{equation}

with the differential emission measure

\begin{equation}
\frac{dEM(h)}{dh}=
n_\mathrm{e}(h)^2 4\pi(R_\star+h)^2 \,.
\label{eq:dEM_dh}
\end{equation}

From Eq.~\ref{eq:Lambdap}--\ref{eq:dEM_dh} and the values of the observed electronic temperatures and emission measures, we derive 
numerically the electronic density at the base of the corona: $n_\mathrm{e,1}(0)=1.9\times10^9$~cm$^{-3}$ and $n_\mathrm{e,2}(0)=6.7\times10^8$~cm$^{-3}$;
Fig.~\ref{Fig:coronal_ne} shows the corresponding profiles of the coronal electronic density.

We obtain the profile of the photon number emitted by the corona for an energy, $E$, and a height, $h$, above the corona:

\begin{eqnarray}
\frac{dN(E_i,h)}{dh} &= & \Delta E_i\times\frac{10^{-14}}{4\pi d^2}\times {\tt vapec}(E_i,kT_1)\times\nonumber \\
		 &  & (\frac{dEM_1(h)}{dh}\nonumber\\
		 &  & +\frac{{\tt vapec}(E_i,kT_2)}{{\tt vapec}(E_i,kT_1)}\times\frac{dEM_2(h)}{dh})\,,
		 \label{Eq:dN_dh} 
\end{eqnarray}

where $\Delta E_i$ is the width of the spectral bin of the ancillary response file at energy $E_i$, and ${\tt vapec}(E_i,kT_j)$ 
is the X-ray photon flux (in unit of photons cm$^{-2}$ s$^{-1}$ keV$^{-1}$) for a plasma temperature of $kT_j$ and a normalization of 1 
(i.e., ${\tt vapec}$ parameter $norm\equiv10^{-14}EM/4\pi d^2=1$).

The ratio of warm-to-cool plasma emissivity in Eq.~\ref{Eq:dN_dh} is plotted versus the energy in Fig.~\ref{Fig:vapec_ratio},
using the energy sampling of the unfolded spectrum. Peaks and dips are due to emission lines from the warm and cool plasma, respectively.
In the 0.25-2~keV energy band, the minimum and maximum values are 0.06 and 262 achieved at 0.57 and 2.00~keV, respectively. For a spectral 
energy resolution of $\Delta E_i=5$~eV, ${\tt vapec}(E,kT_2)/{\tt vapec}(E,kT_1)$ is equal to 0.45 at $E=E_\mathrm{mean}=0.7$~keV.
Figure~\ref{Fig:coronal_dN_dh} shows the resulting normalized profile of the photons emitted by the corona at $E_\mathrm{mean}$ 
obtained using Eq.~\ref{Eq:dN_dh}. Figure~\ref{Fig:coronal_emission_profile} shows the corresponding emission intensity integrated 
along the line of sight versus $R$, the distance from the stellar center. The emission intensity is maximum at the stellar limb 
and decreases from it with the distance, more or less slowly depending of the small or large contribution, respectively, of the warm plasma 
compared to the cool plasma to the coronal emission. Therefore, the size of the coronal emission varies with energy: the smallest and 
largest sizes are observed at 0.57 and 2.00~keV, respectively.

We define the quiescent corona has layers of constant, isotropic, and optically thin emission, above an optically thick sphere of radius 
$R_\star$ (Fig.~\ref{Fig:coronal_dN_dh}). Since ordinary bremsstrahlung emission from plasma shows no intrinsic polarization 
due to the random motions of electrons \citep{Dolan1967}, we consider only unpolarized X-ray photons from the stellar corona. 

\subsection{Model of the HD~189733b atmosphere}
\label{Model:Atmosphere}

The physical properties of the upper atmosphere of HD~189733b can be constrained with transmission spectroscopy of HD~189733A 
\citep[for a review of exoplanetary atmospheres see][]{Madhusudhan2014}. Assuming that the transmission spectrum of HD~189733A from optical 
to near-infrared is dominated by Rayleigh scattering, \cite{Lecavelier2008} derive an atmospheric temperature of $1340\pm150$~K 
at 0.1564~$R_\star$. They prefer scattering by condensates of Enstatite (MgSiO$_3$, with radius between $\sim10^{-2}$ 
and $\sim10^{-1}$~$\mu$m) rather than by molecular hydrogen. Assuming solar abundances, they derive at 0.1564~$R_\star$ for particle 
sizes of about $10^{-2}$ -- $10^{-1}$~$\mu$m, a pressure of $2\times10^{-3}$--$2\times10^{-6}$~bar and a particle density of
$1.1\times10^{16}$--$1.1\times10^{13}$~cm$^{-3}$. The revised transmission spectrum across the entire visible and infrared range 
is well dominated by Rayleigh scattering, which is interpreted as the signature of a haze of condensed grains extending over more 
than two orders of magnitude in pressure, with a possible gradient of grain sizes in the atmosphere due to sedimentation processes 
\citep{Pont2013}.

The gas of the HD~189733b atmosphere is photoionized by the UV emission of HD~189733A \citep[e.g.,][]{Sanz2011} and cools by 
radiation from collisionally excited atomic hydrogen, which leads similarly to H\,{\sc ii} regions to a temperature of about 
$10,000$~K. This high temperature produces  a (slow) evaporative-wind \citep{Yelle2004,Murray2009}. 

\begin{figure}[!t]
  \centering
  \includegraphics[width=0.99\columnwidth]{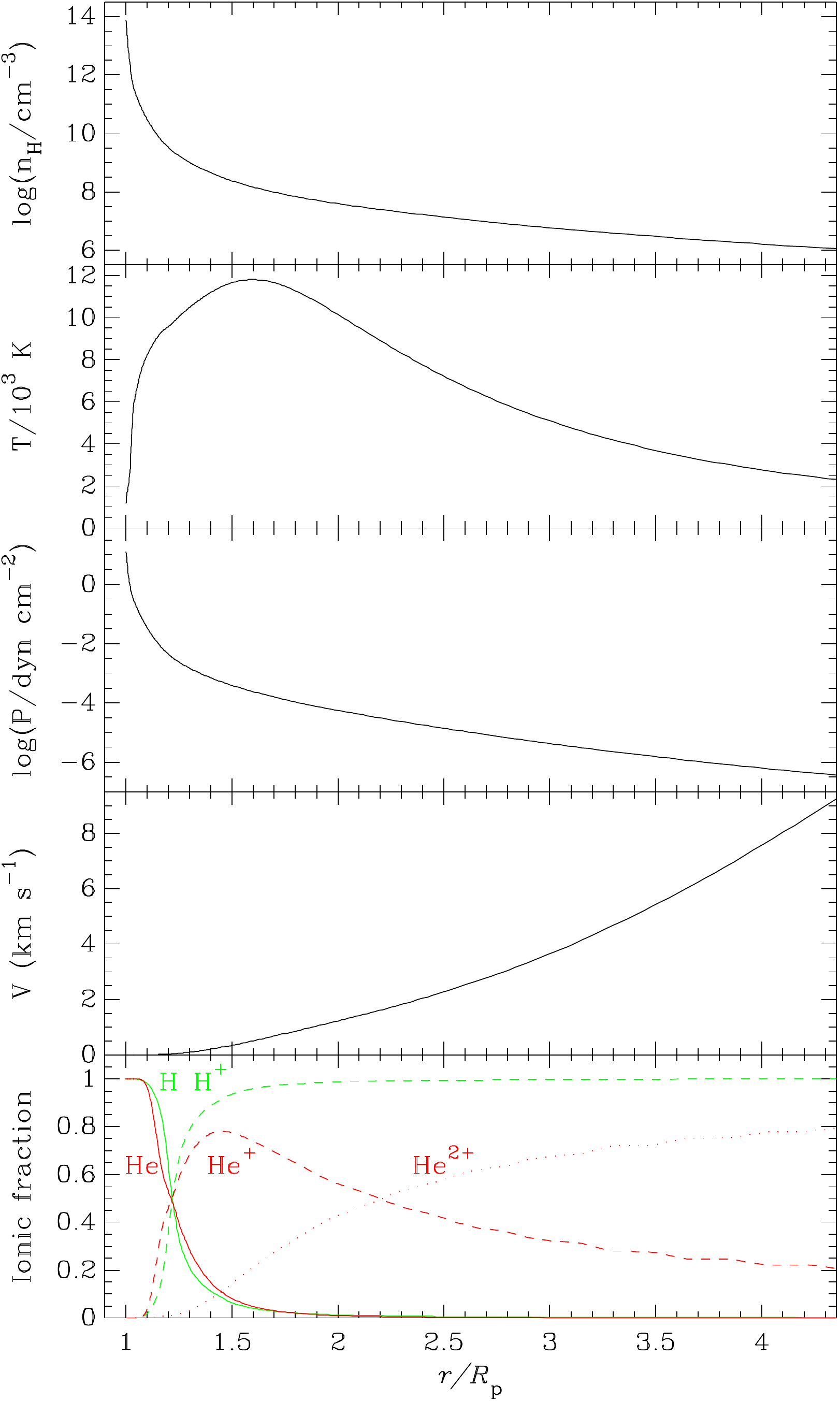} 
  \caption{Atmospheric model of HD~189733b of \citet{Salz2016}
	   used in our X-ray simulation. From top to bottom panels, 
	   the total Hydrogen number density, temperature, pressure, 
	   velocity, and Hydrogen (green) and Helium (red) ionic fractions
	   of the planetary atmosphere are represented versus 
           the distance from the planetary center,  
           ranging from the planetary radius to the Roche's radius.}
  \label{Fig:Model} 
\end{figure}

\citet{Huitson2012} detected an upper atmospheric heating of HD~189733b from HST sodium observations, i.e., evidence of a 
thermosphere. \citet{Salz2015} obtained a 1D, spherically symmetric hydrodynamic simulations of the escaping atmosphere of HD~189733b, 
predicting gas velocity of $\sim$9~km~s$^{-1}$ at the Roche's radius, $R_\mathrm{Roche}=4.35~R_\mathrm{p}$ (Fig.~\ref{Fig:Model}), 
{by coupling a detailed photoionization and plasma simulation code with a general MHD code, and assuming only atomic Hydrogen an Helium 
in the atmosphere}. Since the resulting temperature-pressure profile (see below Fig.~\ref{Fig:P_n_vs_T}) is consistent with the rise 
of temperature with the altitude observed by \citet{Huitson2012}, we adopt the \citet{Salz2015}'s profiles to describe 
for $r \geq R_\mathrm{p}$ the thermosphere of HD~189733b.

\citet{Line2014} showed from secondary eclipse spectroscopy that for $R_\mathrm{p} \geq r \geq r_\mathrm{in}$, 
corresponding to an atmospheric pressure increasing from $p(R_\mathrm{p})\sim$10 to $p(r_\mathrm{in})\sim1000$~dyn~cm$^{-2}$, 
the lower atmosphere of HD~189733b is isothermal, with an equilibrium temperature of $T_\mathrm{eq}\sim1200$~K. Therefore, we 
use the hydrostatic equilibrium to extent the pressure and density profiles into this lower atmospheric layer. The pressure scale 
height of this neutral atmospheric layer is:

\begin{equation}
\lambda_\mathrm{p}^\mathrm{atm}(T_\mathrm{eq})=\frac{kT\mathrm{eq}}{\mu(R_\mathrm{p}) 
m_\mathrm{H} g_\mathrm{p}}
\end{equation}

where $g_\mathrm{p}=GM_\mathrm{p}/R_\mathrm{p}^2$ is the planetary gravitational field, and $T_\mathrm{eq}=1190$~K and 
$\mu(R_\mathrm{p})=1.28015$ \citep{Salz2016}, which leads to $\lambda_\mathrm{p}^\mathrm{atm}(T_\mathrm{eq})=0.0043~R_\mathrm{p}$ 
(i.e., 350~km). The pressure and density follow the same exponential profile:

\begin{eqnarray}
  p(r) 			& = & p(R_\mathrm{p})\times f(r)\\
  n_\mathrm{H}(r) 	& = & n_\mathrm{H}(R_\mathrm{p})\times f(r)
\end{eqnarray}

where

\begin{equation}
  f(r)=\exp{\!\left\{-\frac{r-R_\mathrm{p}}
  {\lambda_\mathrm{p}^\mathrm{atm}(T_\mathrm{eq})}\right\}}
\end{equation}

when neglecting the variation of the planetary gravitional field with the depth below the planetary radius, and 
$p(R_\mathrm{p})=12.7$~dyn~cm$^{-2}$ and $n_\mathrm{H}(R_\mathrm{p})=7.7\times10^{13}$~cm$^{-3}$ \citep{Salz2016}. Therefore, 
$r_\mathrm{in}=0.981R_\mathrm{p}$ and $n_\mathrm{H}(r_\mathrm{in})=6.0\times10^{15}$~cm$^{-3}$.

\subsection{Analytic computation of the transmitted flux}
\label{Model:Transit}

\subsubsection{X-ray absorption radius}
The planetary absorption radius during transit is defined as the distance in the plane of the sky from the planetary center where the optical 
depth through the atmosphere along the line-of-sight, $\tau$, is equal to unity. The total photon extinction cross-section in X-rays is the sum of
photoelectric absorption and scattering cross-sections. The latter is the sum of coherent (Rayleigh) and incoherent (Compton) scattering by 
bound electrons in atoms/molecules \citep{Hubbell1975}. For comparison, we compute these cross-section components for an atmosphere of 
neutral Hydrogen and Helium with solar abundances \citep{Asplund2009}\footnote{The largest solar abundances relative to hydrogen 
in \cite{Asplund2009} are: $A_\mathrm{H}=1$, $A_\mathrm{He}=8.51\times10^{-2}$, $A_\mathrm{O}=4.9\times10^{-4}$, $A_\mathrm{C}=2.69\times10^{-4}$,
$A_\mathrm{Ne}=8.51\times10^{-5}$, $A_\mathrm{N}=6.76\times10^{-5}$, $A_\mathrm{Mg}=3.98\times10^{-5}$, $A_\mathrm{Si}=3.24\times10^{-5}$, 
and $A_\mathrm{Fe}=3.16\times10^{-5}$.} with the {\tt XCOM} \citep{Berger1987}\footnote{The XCOM: Photon Cross Section Database 
(version 1.5; \citealt{Berger2010}) is available at \href{http://physics.nist.gov/PhysRefData/Xcom/Text/intro.html}{
http://physics.nist.gov\-/PhysRefData/Xcom/Text/intro.html} for the energy range of 1~keV--100~GeV.}. At 1~keV the total photon 
cross-section is 46.5~barn/H~atom, composed at 98.1\% by the photoelectric absorption (45.6~barn/H~atom), this proportion decreases 
with higher energy. The scattering cross-section at 1~keV is dominated by the coherent (Rayleigh) process with 0.8~barn/H~atom compared 
to 0.1~barn/H~atom for the incoherent (Compton) scattering. Therefore, we will neglect the scattering cross-sections in our analytic computation
at 0.7~keV.

\begin{figure}[!t]
  \centering
  \includegraphics[width=0.99\columnwidth]{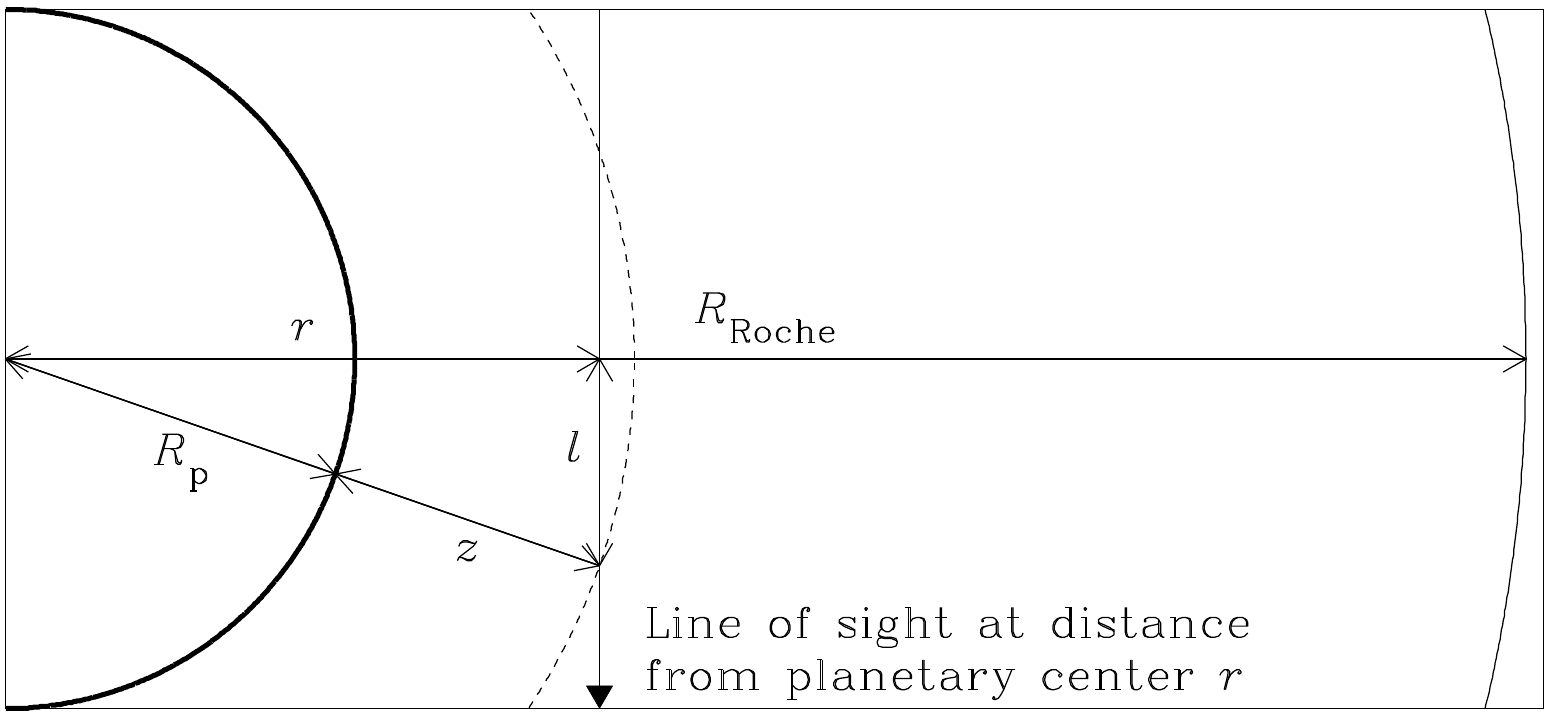} 
  \caption{Definition of the geometrical parameters of the 
  planetary atmosphere.}
  \label{Fig:notations} 
\end{figure}

The simplified solution of the absorption radius proposed by \cite{Fortney2005} is only valid for a geometrically thin atmosphere and can not 
be used {\it a priori} here due to the extended evaporative-wind. From the geometrical parameters introduced in Fig.~\ref{Fig:notations},
the optical depth at energy $E$ through the planetary atmosphere along the line-of-sight is computed versus the distance from the planetary 
center, $r$, as follows:

\begin{equation}
\frac{\tau{_\mathrm{X}(r,E,Z)}}{2}=
\!\!\int_0^L \!\!\!\sigma_\mathrm{X}(\sqrt{r^2+l^2},E,Z) \, n_\mathrm{H}(\sqrt{r^2+l^2})\,dl
\label{tau}
\end{equation}    

with $L=\sqrt{R_\mathrm{Roche}^2-r^2}$, $n_\mathrm{H}$ the total Hydrogen number density, and $\sigma_\mathrm{X}(r,E,Z)$
the photoelectric cross-section per Hydrogen atom, where $Z$ is the metallicity.

The photoelectric cross-section at the distance $r$ from the planetary center and energy $E$ is computed as:

\begin{eqnarray}
\sigma_\mathrm{X}(r,E,Z)&=&f_\mathrm{H}(r)\,\sigma_{\mathrm{X,H}}(E)\nonumber\\
			&+&  A_\mathrm{He}\left(f_\mathrm{He}(r)\,\sigma_{\mathrm{X,He}}(E)+f_\mathrm{He^+}(r)\,\sigma_{\mathrm{X,He^+}}(E)\right)\nonumber\\
			&+&Z \sum_\mathrm{k \notin [H,He]} \!\!A_k\,\sigma_{\mathrm{X,}k}(E)
\label{Eq:depth}
\end{eqnarray}

with $f_\mathrm{H}$, $f_\mathrm{He}$, and $f_\mathrm{He^+}$ the ionic fractions (see bottom panel of Fig.~\ref{Fig:Model});
$A_i$ the abundance in number of the neutral element $i$ relative to the hydrogen \citep{Asplund2009} and $\sigma_{\mathrm{X,}i}$ 
the photoelectric cross-section of the neutral or ionized element $i$ relative to the hydrogen \citep{Verner1995}. Since 
the dust grains in the planetary atmosphere are optically thin to X-rays due to their small sizes ($\ll1$~$\mu$m), there is no 
self-blanketing effect, i.e., no decrease of the X-ray cross-section due to some element in the solid phase \citep{Bethell2011}.

\begin{figure}[!t]
  \includegraphics[width=0.99\columnwidth]{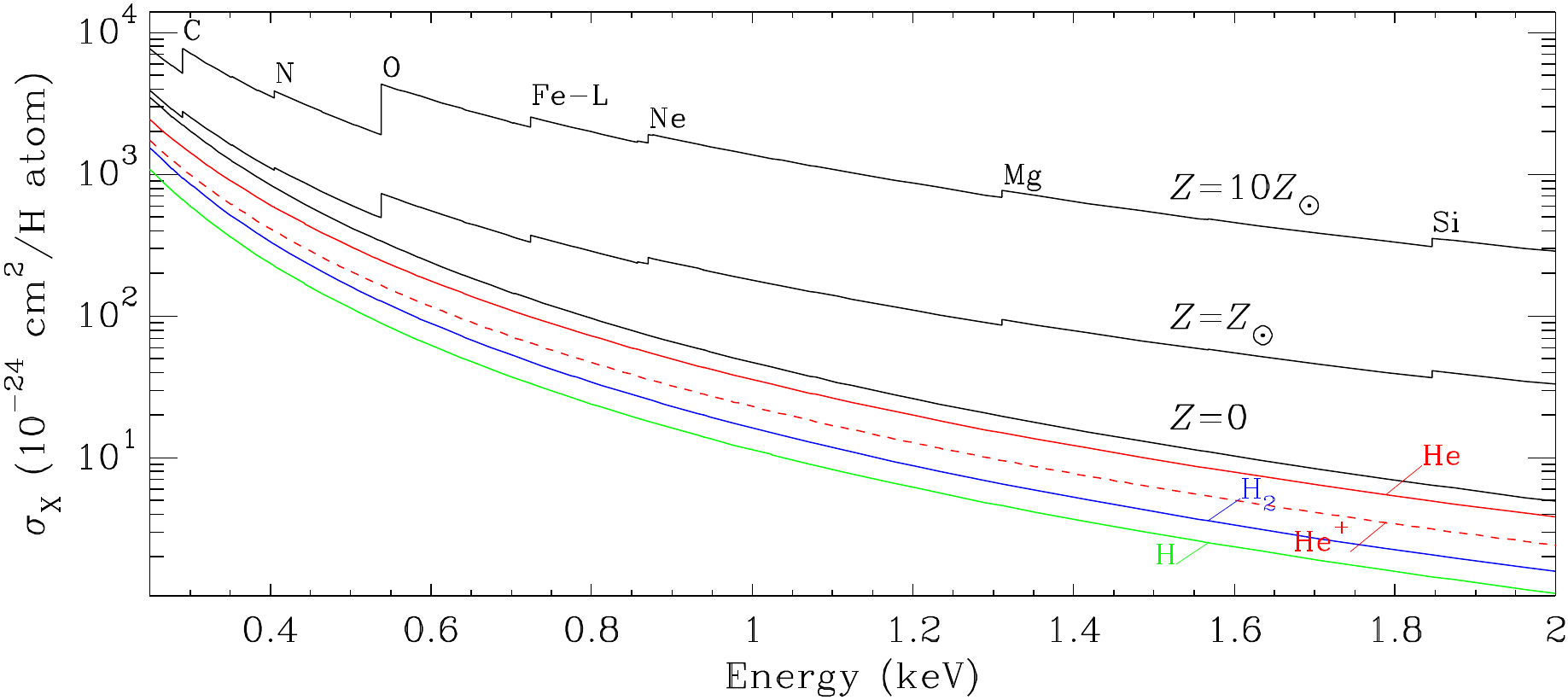}
  \caption{Photon cross-section of photoelectric absorption in X-rays 
	  versus metallicity. The green and blue solid-lines
	  are the X-ray cross-sections of atomic and molecular 
	  Hydrogen, respectively. The solid and dashed red-lines
	  are the X-ray cross-sections of neutral and one-times ionized 
	  Helium, respectively. The black solid lines are 
	  the X-ray cross-section for neutral atomic Hydrogen 
	  and Helium with no metals, solar metallicity and ten times 
	  solar metallicity (no metallic ions). The main 
	  ionization-edges of the inner electronic-shells are 
	  labeled with element names and shell name except for K.}
  \label{Fig:sigma}
\medskip
  \centering
  \includegraphics[width=0.99\columnwidth]{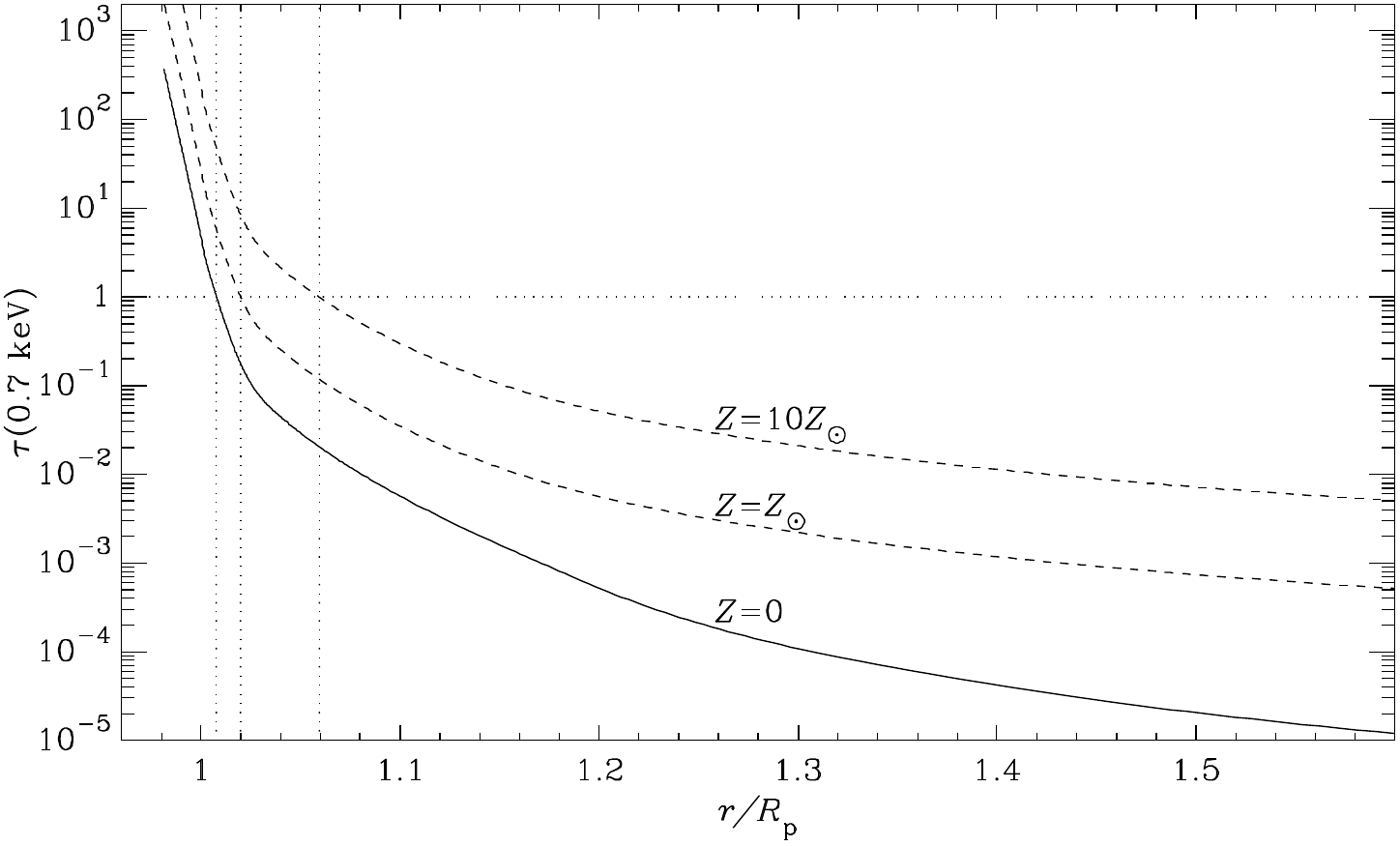}
  \caption{Optical depth through the planetary atmosphere along the line-of-sight
	  at $E_\mathrm{mean}=0.7$~keV versus the distance to the planet 
	  center. The planetary atmosphere is composed of
	  atomic H, He and He$^+$ with no metals, solar 
	  metallicity and ten times solar metallicity (no 
	  metallic ions). The vertical dotted lines indicate the 
	  corresponding X-ray absorption radius.}
  \label{Fig:tau}
\end{figure}

Following \citet{Salz2016}, we will only consider in our simulation atomic Hydrogen and Helium photoelectric 
cross-sections (Fig.~\ref{Fig:sigma}). Since $\sigma_{\mathrm{X,H_2}} = 2.85\,\sigma_\mathrm{X,H}$ \citep{Wilms2000} 
and $A_\mathrm{H_2}=0.5$, the contribution of the molecular Hydrogen in the low atmosphere to an increase of the 
photoelectric cross-section is small. However, metal abundances has a strong impact on the photoelectric absorption.
Since ions (e.g., He$^+$) have lower photoelectric cross-section than neutral elements (fully ionized elements 
have obviously a null photoelectric cross-section), the ionized escaping atmosphere is more transparent to X-rays 
than the lower atmosphere.

\begin{figure}[!t]
\centering
  \includegraphics[width=0.99\columnwidth]{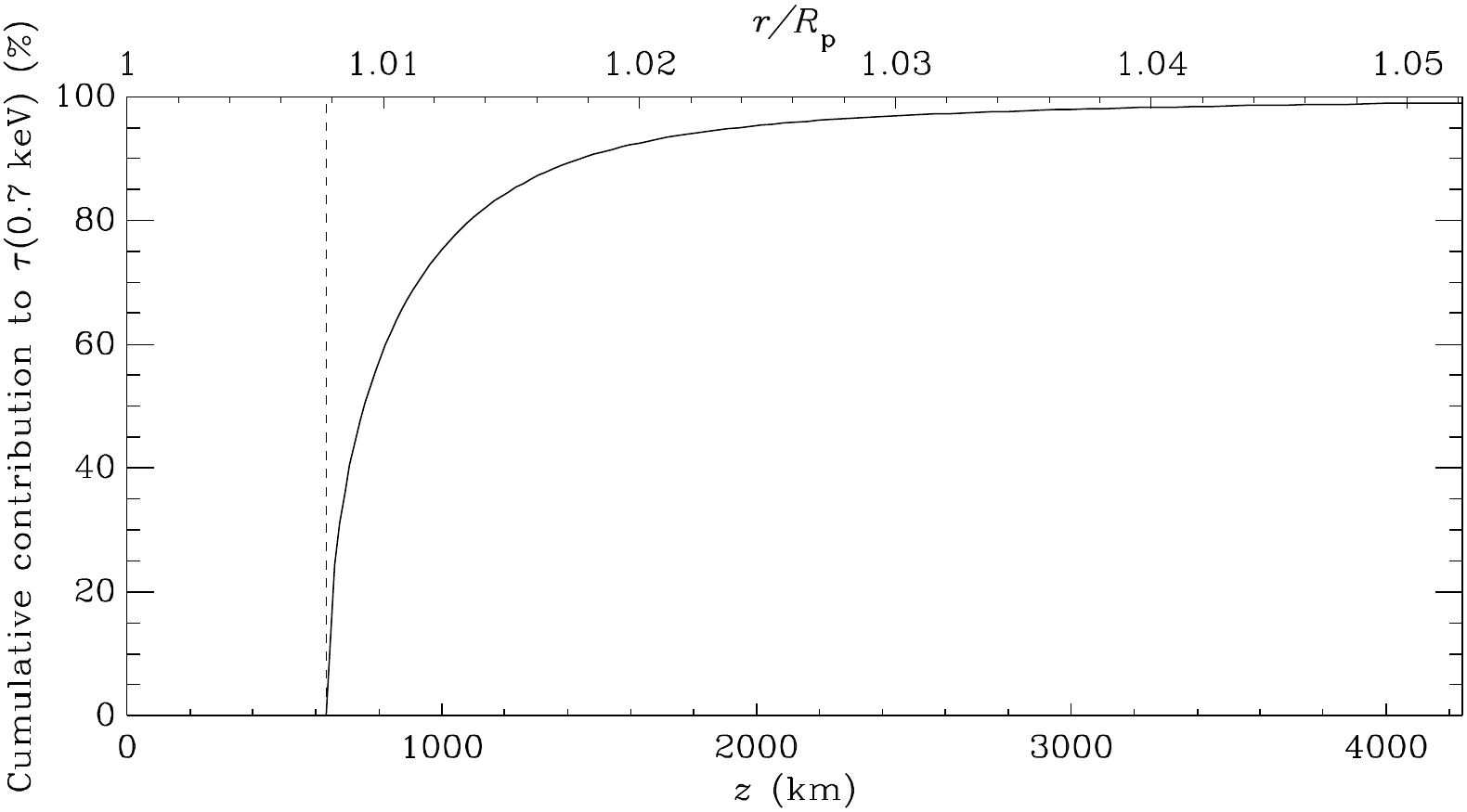}
  \caption{Cumulative contribution of the atmosphere at elevation $z$
	  above the planetary radius to the optical depth at the X-ray absorption 
	  radius at 0.7~keV (vertical dashed line). The maximum elevation 
	  value on the x-axis corresponds to the atmospheric layer contributing 
	  for 99\% to the optical depth.}
   \label{Fig:tau_vs_h}
\medskip
  \includegraphics[width=0.99\columnwidth]{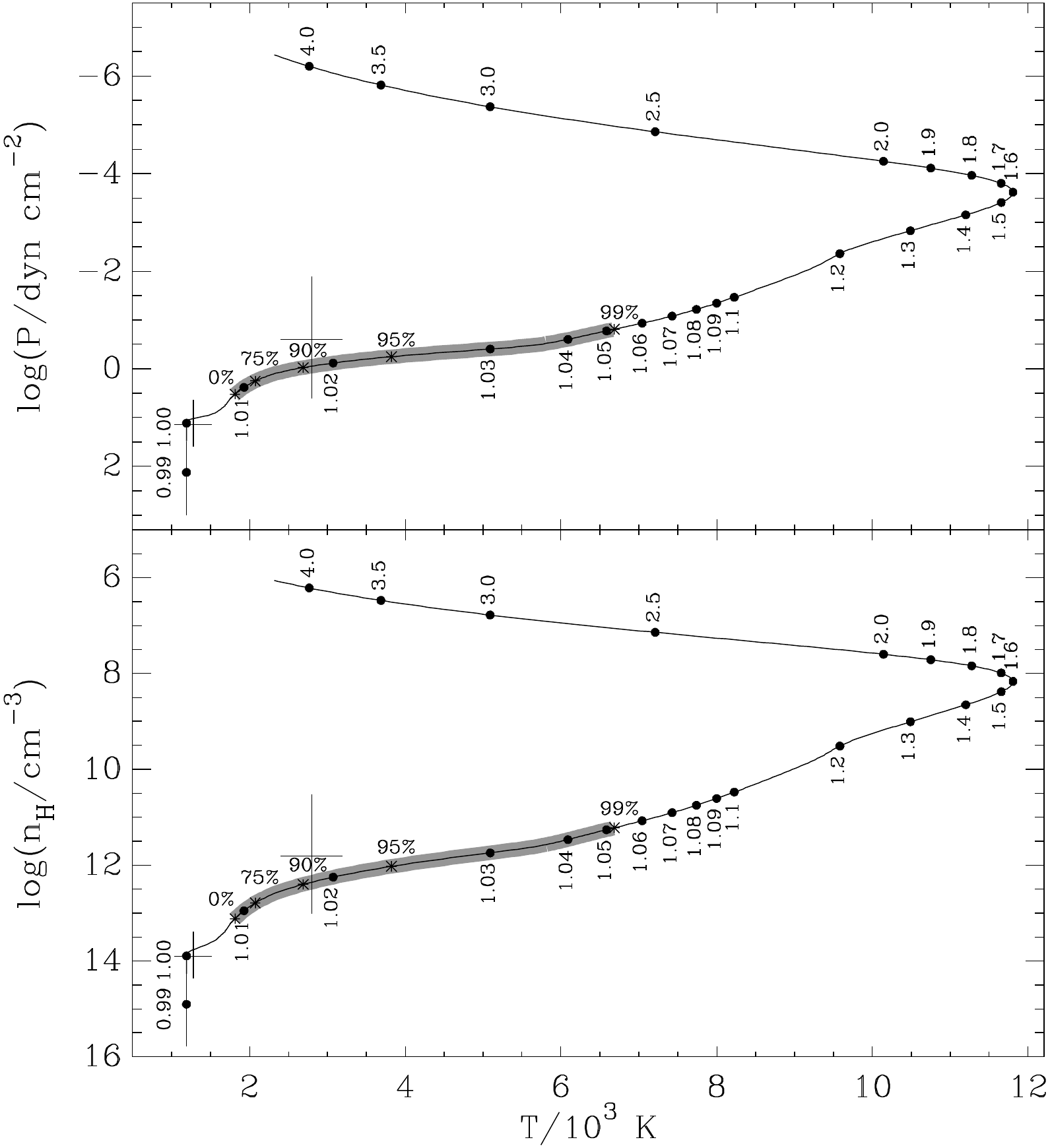}
  \caption{Atmospheric structure of HD~189733b (adapted from 
	  \citealt{Salz2016}). The top and bottom panels are the 
	  pressure-temperature and the density-temperature profile,
	  respectively. The crosses are the HST Sodium observations 
	  of \citet{Huitson2012}. The vertical labels give 
	  the distance from the planetary center in
	  planetary radius. The thick grey 
	  region with percent-labeled asterisks indicates 
	  the contribution of the atmospheric layers
	  to the absorption at the X-ray absorption 
	  radius at 0.7~keV (Fig.~\ref{Fig:tau_vs_h}).}
   \label{Fig:P_n_vs_T}
\end{figure}

Figure~\ref{Fig:tau} shows the variation of the optical depth (Eq.~\ref{Eq:depth}) at 0.7~keV versus 
the distance to the planetary center. We find that the planetary absorption radius at 0.7~keV, $R_\mathrm{X}(\mathrm{0.7~keV})$ 
is 1.008~$R_\mathrm{p}$ (corresponding to 0.156~$R_\star$), i.e., located at $z=$634~km
above the planetary radius. Figure~\ref{Fig:tau_vs_h} shows that 99\% of the absorption at 
$R_\mathrm{X}(\mathrm{0.7~keV})$ is produced by the upper ($z$=634--4243) atmospheric layers that intersect 
the line of sight (e.g., the dashed circle in Fig.~\ref{Fig:notations}). In this absorbing atmospheric 
shell with a width of 3,609~km the total Hydrogen number density and temperature ranges are 
$ 1.1\times10^{13}$--$ 1.2\times10^{11}$~cm$^{-3}$ and 1870--7004~K, respectively 
(Fig.~\ref{Fig:P_n_vs_T}). The physical properties of the atmospheric layers that contribute to the
bulk of the X-ray absorption are constrained by the optical measurements of \cite{Huitson2012}.

Assuming that adding metals in the atmosphere would not change dramatically the density-temperature profile 
(see, however, the discussion in \citealt{Salz2016}), we estimate that $R_\mathrm{X}(\mathrm{0.7~keV})$ increases 
from 1.008 to 1.059~$R_\mathrm{p}$ (or equivalently from 0.156 to 0.164~$R_\star$) when the neutral 
metals increase from zero to ten times the solar metallicity (Fig.~\ref{Fig:tau}). 

\subsubsection{Transmitted flux at 0.7~keV during the planetary transit on the stellar corona}
\label{Sect:transit}
\begin{figure}[!t]
  \centering
  \includegraphics[width=\columnwidth]{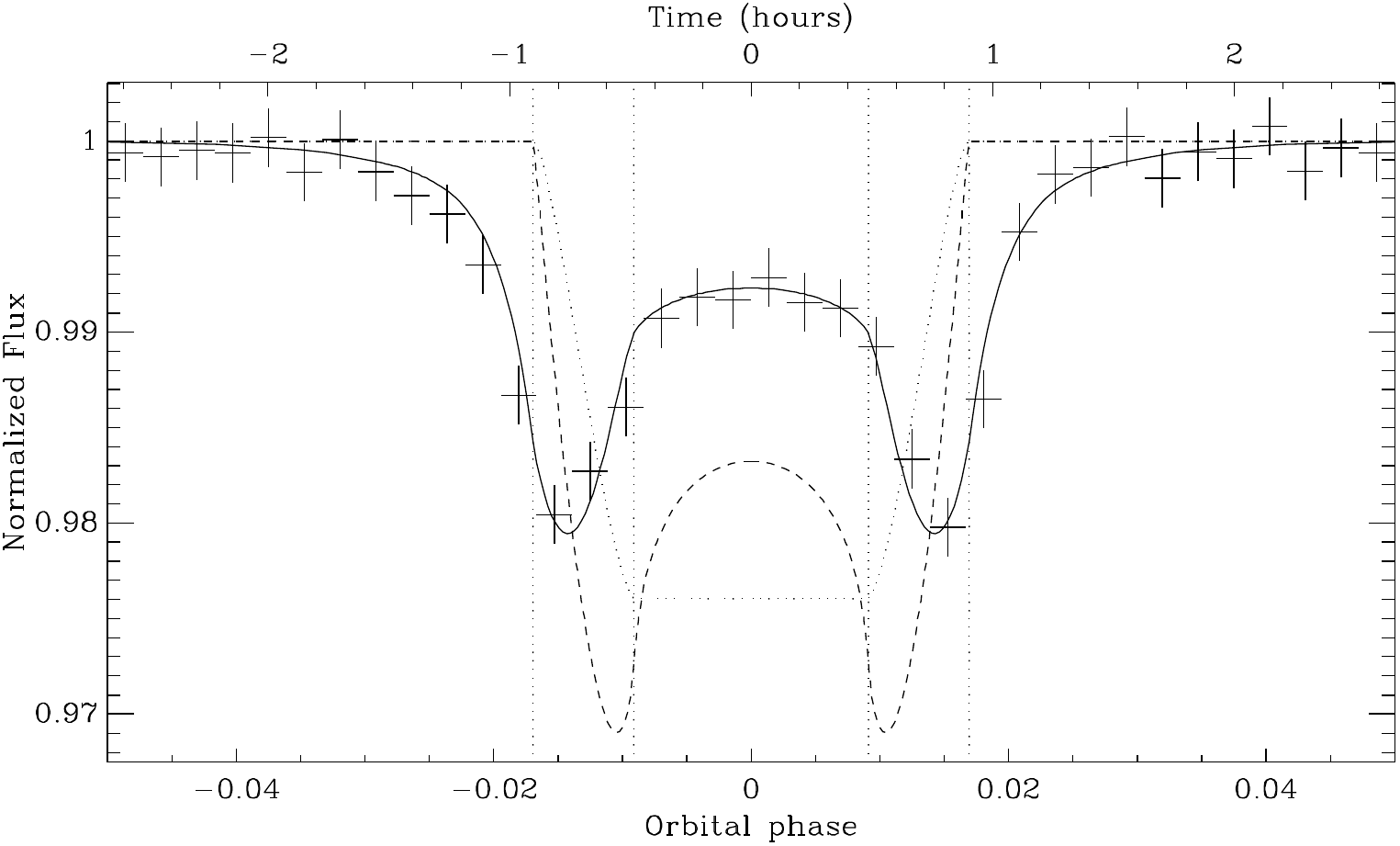}
  \caption{Light curves of the planetary transits of HD~189733b.
	  The black solid line is the coronal transit
	  at 0.7~keV computed for $R_\mathrm{X}(\mathrm{0.7~keV})=1.008R_\mathrm{p}$ 
	  (corresponding to a planetary atmosphere of H, He, 
	  and He$^+$). For comparison, the vertical dotted
	  lines indicated the first, second, third, and fourth contacts
	  of the photospheric transit in the optical. The 
	  dotted and dashed lines are the photospheric transit 
	  (uniformly emitting disk, e.g., \citealt{Mandel2002})  
	  and the chromospheric transit \citep{Schlawin2010}, respectively.
	  The cross data are the coronal transit signature obtained 
	  by our simulation at 0.7~keV (see Fig.~\ref{Fig:Results}).}
  \label{Fig:Transit}
\end{figure}

For comparison purpose, we compute the phases of the first, second, third and fourth contacts of the primary 
eclipse in the optical\footnote{The formulae for the phases of the first and second contacts are respectively:\\
$\phi_\mathrm{1}=-\arcsin{\!(\sqrt{(R_\star+R_\mathrm{p})^2-a^2\cos{i}^2}/a\sin{i}})/2\pi$ and\\
$\phi_\mathrm{2}=-\arcsin{\!(\sqrt{(R_\star-R_\mathrm{p})^2-a^2\cos{i}^2}/a\sin{i}})/2\pi$.}: 
$\phi_1=-\phi_4=-0.0169$ and $\phi_2=-\phi_3=-0.0091$ (vertical dotted lines in Fig.~\ref{Fig:Transit}).
The corresponding photospheric transit of a uniformly emitting disk \citep[e.g.,][]{Mandel2002} has a maximum transit 
depth of $(R_\mathrm{p}/R_\star)^2=2.39\%$ (dotted line in Fig.~\ref{Fig:Transit}).

The chromospheric transit was computed analytically by \cite{Schlawin2010}, assuming an optically thin 
and geometrically thin shell of emission (zero thickness). It is characterized by a W-profile with a maximum depth 
of approximately $\delta_\mathrm{max,approx}^\mathrm{chrom}=0.5 (R_\mathrm{p}/R_\star)^{3/2}$, i.e., 
$0.5 (R_\mathrm{p}/R_\star)^{-1/2}$ times deeper than the maximum transit depth of a uniformly emitting disk.
These two minimums occur slightly before and after the second and third contacts, respectively, with a depth 
here of exactly $\delta_\mathrm{max}^\mathrm{chrom}=3.10\%$ (dashed line in Fig.~\ref{Fig:Transit}). This 
characteristic shape is due to the strong limb brightening that occurs since the zero-thickness chromosphere 
has its largest column density at the disk's edge.

We compute numerically the normalized light curve of the planetary transit on the stellar corona from 
the coronal emission profile integrated along the line of sight, $I$ (Fig.~\ref{Fig:coronal_emission_profile}), 
and the optical depth through the planetary atmosphere, $\tau$ (Fig.~\ref{Fig:tau}), using the formula:

\begin{equation}
I_\mathrm{transit}(t) = \frac{\int_{-\infty}^{+\infty}\int_{-\infty}^{+\infty} I(\sqrt{X^2+Y^2})\,e^{-\tau\left(\frac{r(X,Y,t)}{R_\mathrm{p}}\right)}dXdY}{\int_{-\infty}^{+\infty}\int_{-\infty}^{+\infty} I(\sqrt{X^2+Y^2})\,dXdY}
\label{Eq:transit}
\end{equation}

where $X$ and $Y$ are the cartesian coordinates in the plane of the sky expressed in stellar radius (Fig.~\ref{Fig:Sketch}), 
and $r(X,Y,t)=\sqrt{(X-X_\mathrm{p}(t))^2+(Y-Y_\mathrm{p}(t))^2}$ is the distance from the position of the planetary center at time $t$, 
$X_\mathrm{p}(t)=a\cos(i)\sin(2\pi t/P)$ and $Y_\mathrm{p}(t)=-a\cos(i)\cos(2\pi t/P)$. We obtain a similar W-profile since the 
coronal emission is also optically thin (solid line in Fig.~\ref{Fig:Transit}). But, the coronal transit is less deep 
and more extended than the chromospheric transit since the corona is extended above the photosphere (green line in 
Fig.~\ref{Fig:coronal_emission_profile}). We find $\delta_\mathrm{max}=2.06\%$ at $\phi=\pm0.0143$, corresponding to a 
temporal shift of $\pm$45.8~min from the transit center. At the transit center the depth is reduced to only 0.77\%.

\subsubsection{Impact of the broad energy-band on the observed transmitted flux}
\label{Sect:broad_band}

\begin{figure}[!t]
  \centering
    \includegraphics[width=\columnwidth]{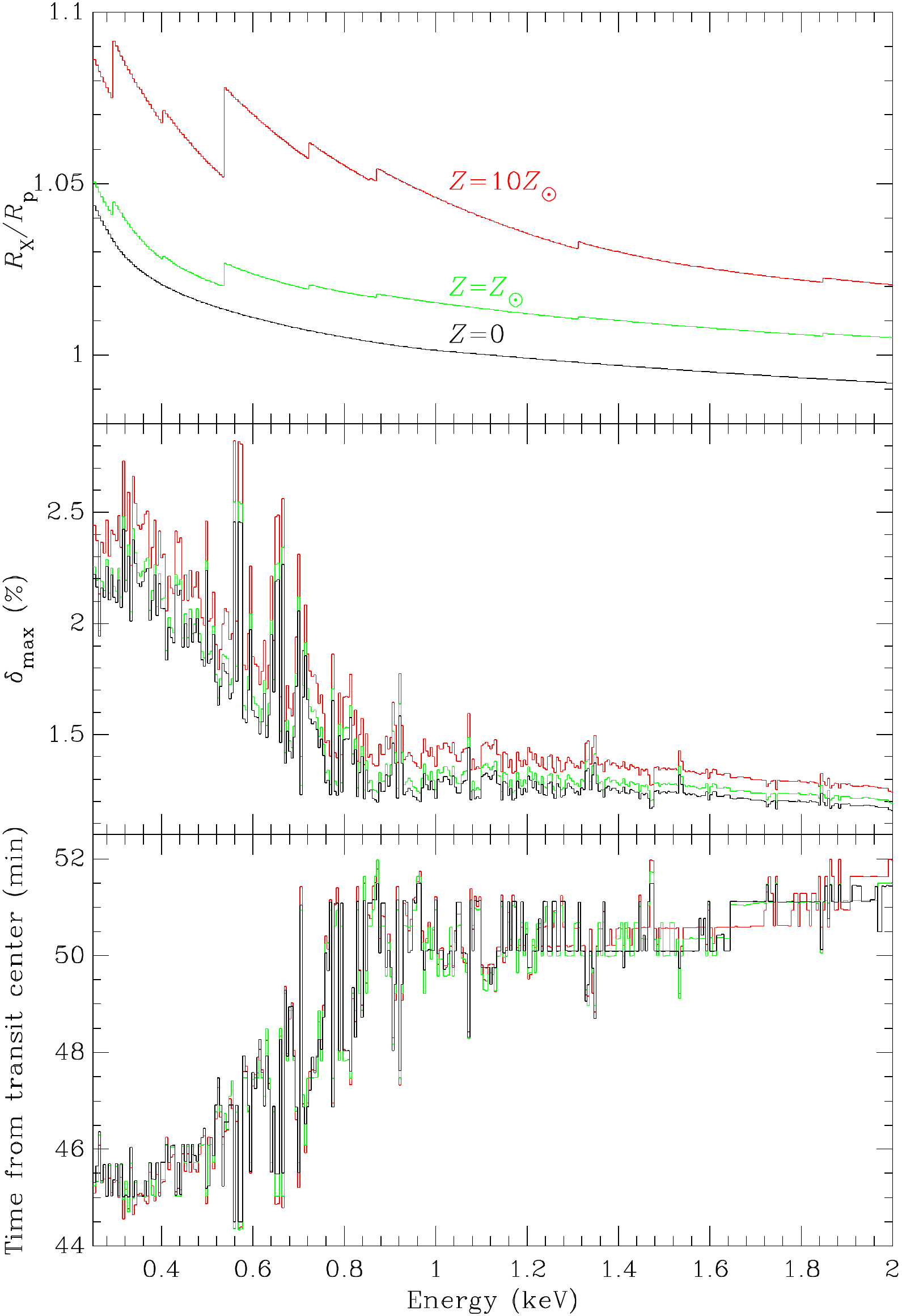}
     \caption{Characteristics of the coronal transit versus X-ray energy. 
	      {\it Top panel:} X-ray absorption radius given in planetary radius.
	      {\it Middle panel:} maximum depth of the coronal transit in percent.
	      {\it Bottom panel:} temporal shift of the maximum depth from the 
	      transit center. Black, green, and red lines are for a planetary 
	      atmosphere composed of atomic H, He, He$^+$ with no metals, solar 
	      metallicity and ten times solar metallicity (no metallic ions), 
	      respectively.}
  \label{Fig:delta_max}
   \includegraphics[width=\columnwidth]{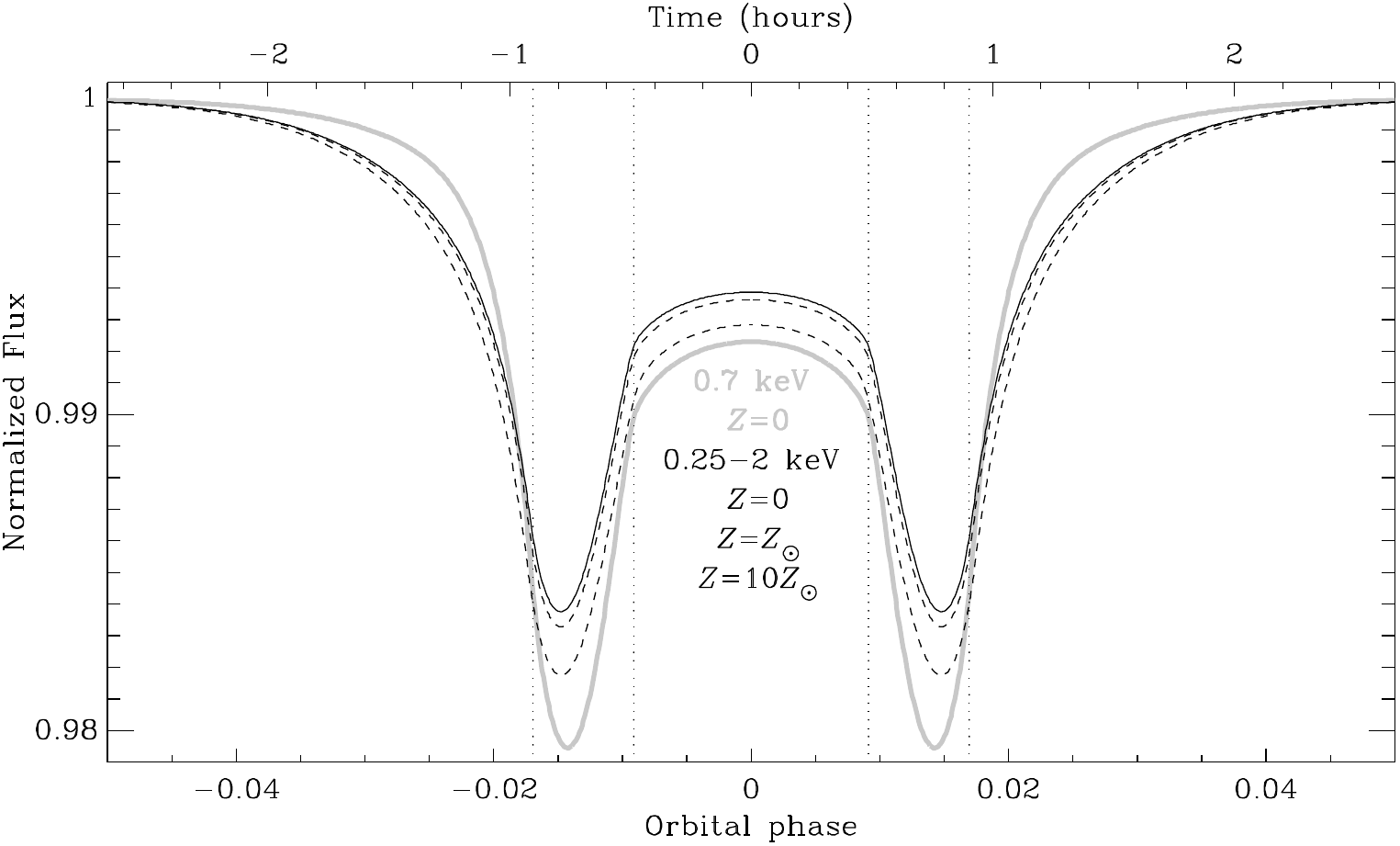}
     \caption{Planetary transit in the 0.25--2.0~keV energy band. For 
	      comparison purpose, the vertical dashed lines are the eclipse 
	      contact in the optical and the grey line is the planetary transit 
	      at 0.7~keV for no metals.}
  \label{Fig:broad_band_transit}
\end{figure}

Due to the limited sensitivity of the current X-ray facilities, any tentative detection of planetary transits 
must use light curves in a broad energy band to increase the signal-to-noise ratio. The extension and 
depth of the coronal transit is modified by the variations with energy of the coronal emission profile, the photon
extinction cross-section (determining the absorption radius), and the observed stellar flux. It is therefore 
important to assess in a broad energy band the effective transit. This effective transit is the weighted
average of the monoenergetic transit on the broad energy-band, using the observed X-ray count rate versus 
energy (i.e., the observed X-ray spectrum; top panel of Fig.~\ref{Fig:pn_spectrum}) as weight.

For a given metallicity, we compute at each grid energy the coronal profile and the absorption radius, 
and the corresponding light curve of the coronal transit. The top panel of Fig.~\ref{Fig:delta_max} shows the 
absorption radius decreasing with the energy, with local jump associated to the ionization edges of metals.
When there is no metals, the X-ray absorption radius becomes lower than the planetary radius --the
lower boundary of \citet{Salz2016}'s simulation-- at energies larger than $\sim$1.1~keV. Therefore, the optical 
depth used to compute the absorption factor in Eq.~\ref{Eq:transit} has also to be estimated inside the lower 
atmosphere, which requires an extension of the density model as we did in Sect.~\ref{Model:Atmosphere}.
This leads for comparison at $r_\mathrm{in}$ to $\tau(Z=0,E=2.0~\mathrm{keV})\sim12$.

The mid and bottom panels of Fig.~\ref{Fig:delta_max} show the maximum depth of the coronal transit and its 
temporal shift from the transit center, respectively, vs.\ energy. Both values are mainly correlated with the
warm-to-cool plasma emissivity ratio (Fig.~\ref{Fig:vapec_ratio}) that controls the extension of the corona.

Then, we compute the weighted average light curve of the coronal transit in the 0.25--2~keV energy band for 
XMM-Newton/pn using the template instrumental spectrum of HD~189733A as weight (Fig.~\ref{Fig:broad_band_transit}). 
We obtain maximum transit depths of 1.62\%, 1.67\%, and 1.82\%, for no metals, and one and ten times solar 
metallicity (no metallic ions), respectively, at $\phi=\pm0.0148$ corresponding to a temporal shift of 
$\pm$47.4~min from the transit center; at the transit center the depth is reduced to 0.61\%, 0.64\%, and 0.72\%, 
respectively.

Therefore, the maximum transit depth in the 0.25--2~keV energy band is little sensitive to the metal 
abundance. Moreover, the transit light curve at 0.7~keV (grey line in Fig.~\ref{Fig:broad_band_transit}) is a 
correct approximation of the transit light curve in the 0.25--2~keV energy band.

\subsection{Monte Carlo radiative transfer}
\label{Model:Code}

To simulate the transit of HD~189733b in front of its host star, we used the radiative transfer code {\sc stokes}. Originally written for 
polarimetric investigations of active galactic nuclei, {\sc stokes} has been extended over the last years to cover a much larger panel of 
astrophysical sources (a detailed list of examples can be found in \citealt{Marin2014}). The code and its characteristics are extensively 
described in \citet{Goosmann2007}, \citet{Marin2012} and \citet{Marin2015b} for the optical/UV part, and in \citet{Marin2015} for the X-ray band, 
so we will simply summarize the principal aspects of {\sc stokes} relevant for our computations.

The code can handle a large panel of emitting, scattering and/or absorbing geometries in a fully three-dimensional environment. Multiple scattering
enables the code to radiatively couple all the different regions of a given model, and a web of virtual detectors arranged in a spherical geometry is used
to record the photon wavelength, intensity, and polarization state, according to the Stokes formalism ($\vec S = (I, Q, U, V)^{\rm T}$).
About 10$^{ 11}$ photons have been sampled for our analysis. The total intensity (flux), polarization degree $P$ (= $\sqrt{Q^2+U^2+V^2}/I$), 
and polarization angle $\psi$ (= $0.5\tan^{-1}(U/Q)$) are computed by averaging all the detected photons, at each line-of-sight in polar and azimuthal 
directions. Mie, Rayleigh, Thomson, Compton and inverse Compton scattering are included from the near-infrared to hard X-ray bands. 

\begin{figure}
  \centering
  \includegraphics[trim = 0mm 5mm 0mm 25mm, clip, width=9.9cm]{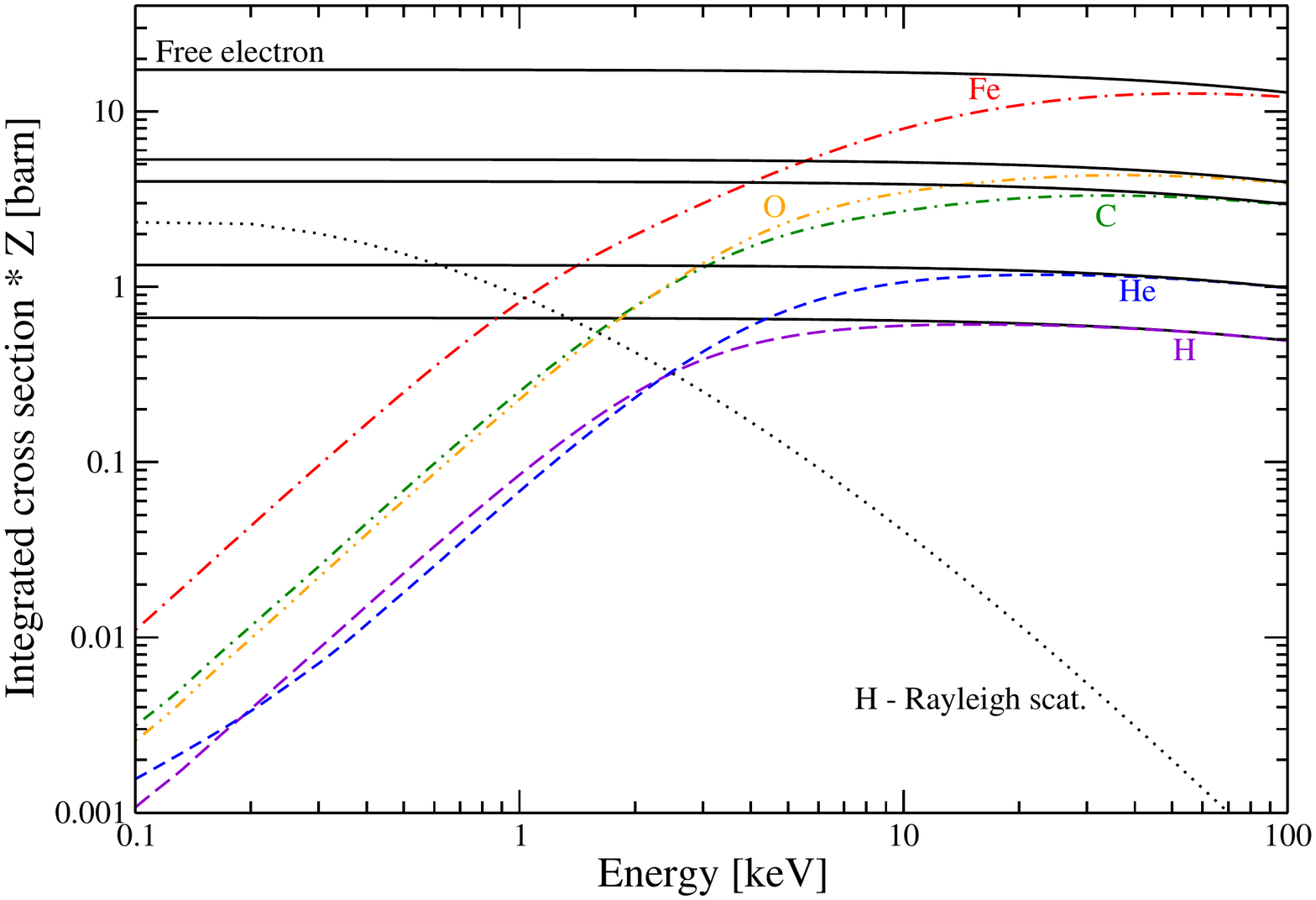} 
  \includegraphics[trim = 0mm 5mm 0mm 25mm, clip, width=9.9cm]{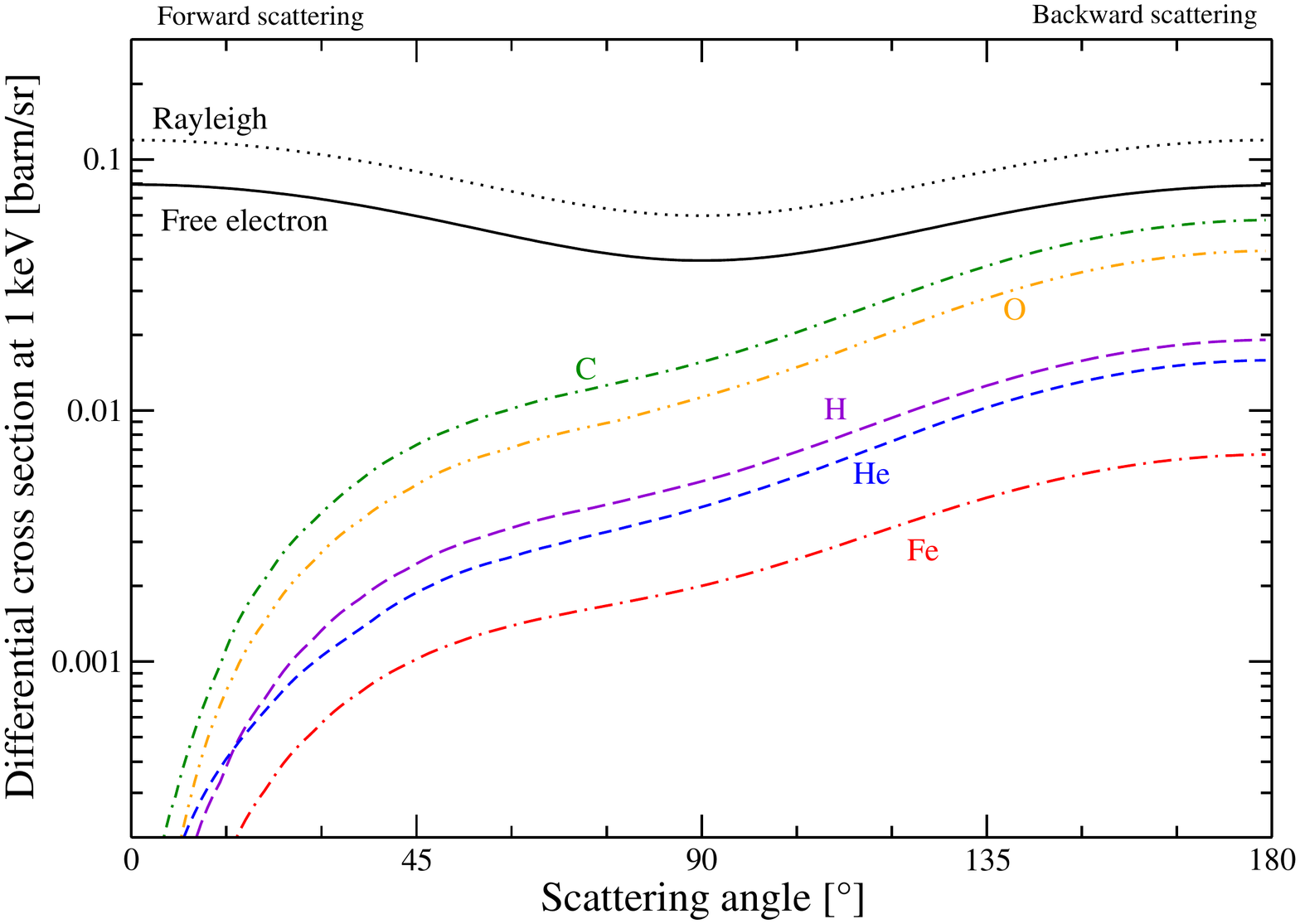} 
  \caption{Elemental photon scattering cross-section versus energy.
	   Top: integrated cross-section for Compton scattering 
	   onto a variety of elements chosen accordingly to \citet{Asplund2009} 
	   for their high abundance. Solid black line: coherent (Compton) 
	   scattering on free electrons (Klein-Nishina cross-section).
	   Incoherent (Compton) scattering on bound electrons: long-dashed violet 
	   line of H, short-dashed blue line of He, dot-short-dashed green line of C, 
	   double-dot-dashed orange line of O, dot-long-dashed red line of Fe.
	   The black dotted line shows coherent (Rayleigh) scattering onto 
	   an hydrogen atom for comparison. Bottom: $E_\mathrm{mean}=0.7$~keV 
	   differential cross section versus scattering angle for the same 
	   free and bound electrons.}
  \label{Fig:ElectronBound} 
\end{figure}

For planetary atmospheres in the X-ray domain, Rayleigh and Compton scattering of photons onto bound electrons will be the major mechanism 
to alter the net polarization, and a visual representation of the integrated and differential scattering cross sections produced by {\sc stokes} 
is presented in Fig.~\ref{Fig:ElectronBound}. Note that for molecular hydrogen the incoherent scattering cross-section is two-times larger 
than for atomic hydrogen \citep{Sunyaev1999}.
 
Photo-ionization and recombination effects, which also influence the resulting polarization by absorption and unpolarized re-emission processes, 
are based on quantum mechanics computations from \citet{Lee1994a} and \citet{Lee1994b}. We use fluorescent yields and energy of the subsequently 
re-emitted Auger photons from the Opacity Project \citep{Cunto1992}, and X-ray photoelectric cross-sections and inner-shell edge energies of \cite{Verner1995}.
Finally, we opt for elemental abundances from \citet{Asplund2009} and ten times solar metallicity to allow direct comparison with \cite{Poppenhaeger2013}.

Since the X-ray scattering by dust grains \citep{Mathis1991} is not yet implemented in {\sc stokes}, we consider only gaseous components. 
We do not consider additional soft X-ray emission from the exoplanet itself induced by the stellar wind charge exchange mechanism 
since it was estimated in the case of the close-in exoplanet HD~209458b to be about five orders of magnitude fainter than the emission 
from the host star \citep{Kislyakova2015}.

We divide our two-zone model with constant-density shells sampling any 0.1-dex variation of the density from a density of 7.697$\times$10$^{13}$~cm$^{-3}$ 
corresponding to a pressure of 12.739~bar, a temperature of 1190.2~K, and a 0.7~keV optical depth of $\sim$4.608.

\section{Photo-polarimetric results}
\label{Results}
 
We present the outcomes of our simulations in Fig~\ref{Fig:Results}. Results are plotted with respect to the orbital phase of HD~189733b,
also converted into transit hours. Each data-point is associated with its 1$\sigma$ Monte Carlo error bars resulting from the simulation. 
The Monte Carlo statistical error is estimated using the standard deviation $\sigma$ that is roughly proportional to $\sqrt{\rm Var(X)}$/$\sqrt{N}$ 
for large $N$, with $N$ the number of photons sampled in the direction of the observer and Var(X) the variance of the estimate.

\subsection{Photometry}
\label{Results:photometry}

\begin{figure*}
  \centering
  \includegraphics[trim = 0mm 140mm 18mm 8mm, clip, width=19.5cm]{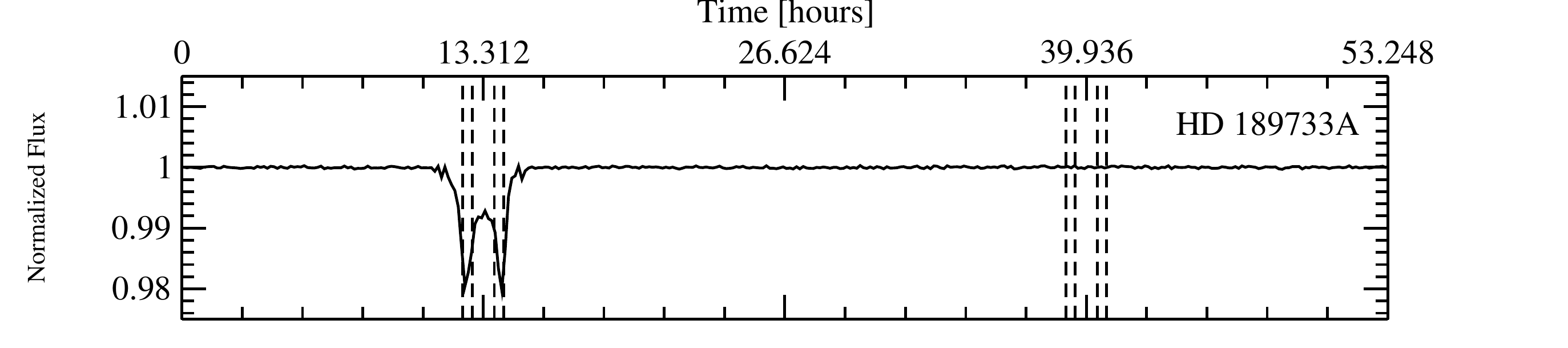} 
  \includegraphics[trim = 0mm 0mm 0mm 8mm, clip, width=20.8cm]{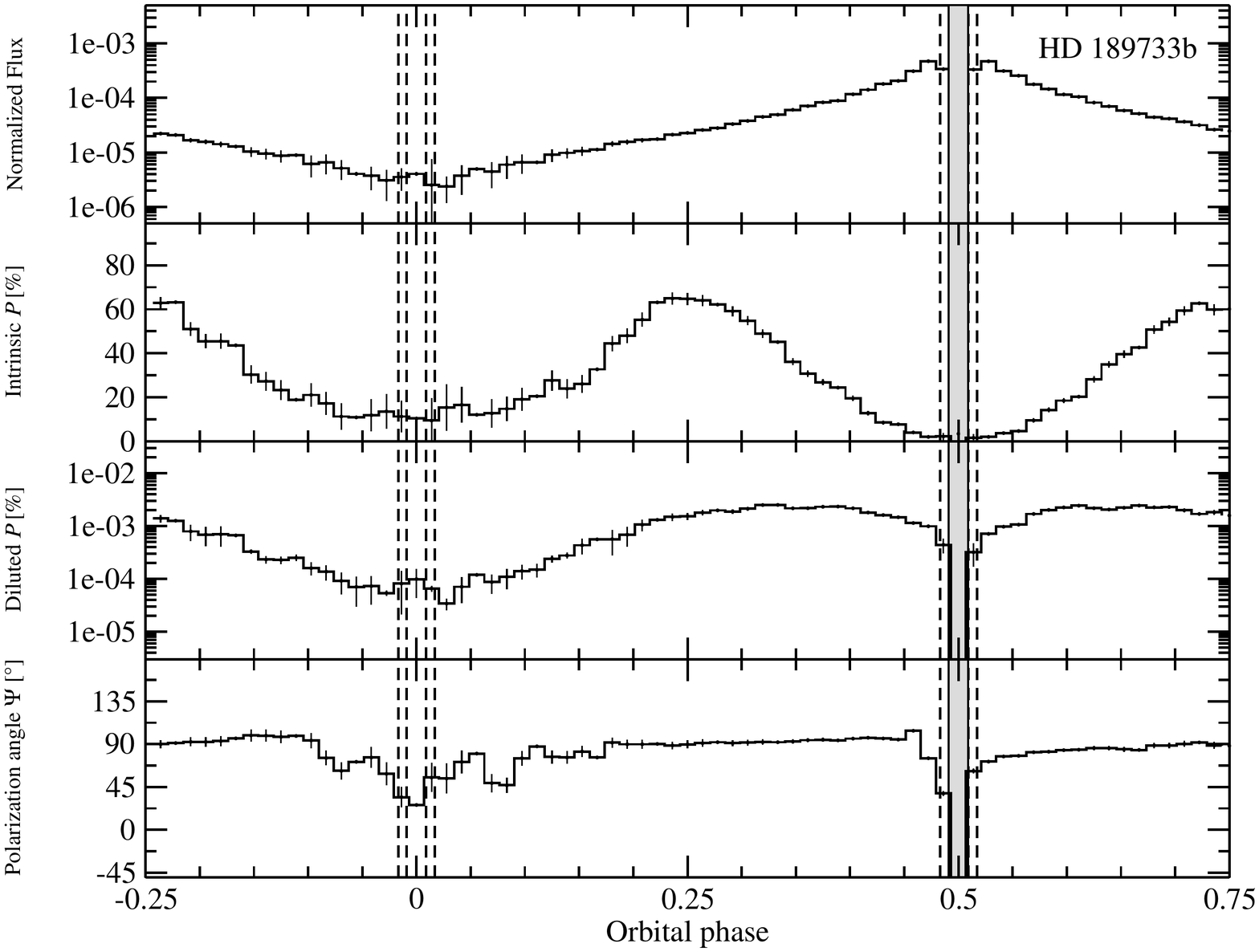} 
  \caption{Photometry at $E_\mathrm{mean}$=0.7~keV of HD~189733A, and reprocessed flux from the surface of HD~189733b (first 
	   two panels). The fluxes are normalized to the initial flux of the star. Polarimetric results
	   are shown in terms of $P$ (polarization percentage) and $\psi$ (polarization position angle, 
	   defined with respect to the vertical axis of the system). The error bars correspond to 
	   1$\sigma$ statistical errors in the Monte Carlo simulation. The vertical dashed lines are 
	   the first, second, third and fourth contacts of the primary (planetary transit) and 
	   secondary eclipses in the optical. The gray box indicates a region where polarization 
	   measurement of HD~189733b is impossible due to the disappearance of the exoplanet behind 
	   its star.}
  \label{Fig:Results} 
\end{figure*}

The top panel of Fig~\ref{Fig:Results} shows the light curve of HD~189733A at 0.7~keV, normalized to unity. The exoplanet 
orbits in front of HD~189733A from orbital phase -0.25 to orbital phase +0.25, with an eclipse of the stellar corona 
(i.e., a planetary transit) occuring between orbital phases -0.0169 and +0.0169. During this planetary transit, the 
0.7~keV flux decreases by 2.00$\pm$0.15~\% between the first and second contacts, as the exoplanet starts to cover the X-ray 
corona, then rises to 1.07$\pm$0.15~\% at orbital phase 0. It then decreases back to 2.02$\pm$0.15~\% between the third and 
fourth contacts before resuming to the non-eclipsed flux, showing no more variation during the remainder of the orbit of HD~189733b. 
A close up representation of the transit is shown in Fig.~\ref{Fig:Transit} for clarity purposes. We find a perfect agreement 
between numerical and simulation results within the Monte-Carlo statistical error of our modeling.

Using the {\sc stokes} code, we isolated photons that are re-emitted/scattered from HD~189733b, and plotted the resulting 
light curve (normalized with respect to the unitary flux of HD~189733A) in Fig~\ref{Fig:Results} (second panel from the top). 
To increase the statistics of the photon-starved results, the phase resolution of the re-emitted/scattered radiation
has been decreased compared to Fig.~\ref{Fig:Transit}. Due to the strong photoelectric absorption processes that prevail 
at soft X-ray energies, the X-ray flux from the exoplanet is found to be three to five orders of magnitude lower than the photon 
flux from the main star. From the HD~189733A flux of $2.9\times 10^{-13~}$~erg~cm$^{-2}$~s$^{-1}$ at 0.7~keV, the flux of HD~189733b 
is about $10^{-16}$ -- $10^{-18}$~erg~cm$^{-2}$~s$^{-1}$, depending on the orbital phase of the planet. The flux reprocessed onto the 
atmospheric layers of HD~189733b strongly depends on the angle between the source, the exoplanet atmosphere and the observer; this 
effect is due to the differential scattering phase function of incoherent scattering, as presented in Fig.~\ref{Fig:ElectronBound} 
(bottom). In the case of Rayleigh scattering, forward and backward reprocessing prevails while, for incoherent scattering onto bound 
electrons, forward and perpendicular scattering are less probable. However, backscattering is very strong and when the gaseous planet 
passes beyond the plane-of-the-sky, its reprocessed flux rises up to a normalized flux of $\sim$~4.7 $\times$ 10$^{-4}$.
Then, HD~189733b disappears behind its star and flux drops to zero.

\subsection{Polarization}
\label{Results:polarization}

The middle panel of Fig~\ref{Fig:Results} is a visualization of the linear continuum polarization degree $P$, ranging from 0 
(unpolarized) to 100~\% (fully polarized). It is the polarization resulting from scattered light only, i.e. not diluted by the unpolarized 
stellar flux. The shape of the polarization curve is very similar to the results obtained by \citet{Stam2004}, who evaluated the spectra 
of visible to near-infrared starlight reflected by Jupiter-like extrasolar planets. As \citet{Stam2004}, we find polarization maximums 
close to orbital phases $\pm0.25$, i.e., near planetary greatest elongations, with $P$ = 64.9$\pm$2.9~\% which is slightly larger
than in infrared. Since the planet's orbital plane is inclined by 4$^{\circ}$, the polarization degree does not drop to zero at zero orbital 
phase as a small fraction of photons scatters onto the top of the atmospheric layers towards the Earth. Hence, we are sure that our model 
produces correct results and extends the near-infrared/optical investigation of \citet{Stam2004} to X-ray energies.

The fourth panel of Fig~\ref{Fig:Results} also shows the star+planet linear polarization, now taking into account polarization
dilution by the unpolarized stellar photons, i.e., $P_\mathrm{diluted}=\sqrt{Q^2+U^2+V^2}/(I_\star+I_\mathrm{p,scatt})$. The overall
shape of the polarization curve, associated with a polarization degree below 0.003~\%, is quite different from the intrinsic $P$. 
In fact, maximum polarization is found when the exoplanet transits behind the plane-of-the-sky, with a local diminution to zero when HD~189733b
is occulted by its primary star. The departure from a purely sinusoidal waveform is again due to the scattering phase function of incoherent
scattering that promotes backscattering. Associated with maximum fluxes from HD~189733b that are not located at maximal planetary elongations,
the net diluted polarization degree peaks at orbital phases +0.35 and +0.65. The highest degrees of $P$ are of the order of 0.0025~$\pm$~0.0002~\%
at 0.7~keV, similarly to the polarization found in the optical band \citep{Carciofi2005}.

Finally, the bottom panel of Fig~\ref{Fig:Results} presents the absolute polarization position angle $\psi$ to be observed with an X-ray polarimeter.
When HD~189733b does not transit on HD~189733A, its polarization angle is equal to 90$^{\circ}$ (the $\vec E$-vector of the radiation is aligned 
with the projected rotational axis of the orbit). However, when the planet is about to transit in front of its primary star, $\psi$ oscillates
around 90$^{\circ}$. The canceling contribution of unpolarized fluxes from the star and the 90$^{\circ}$ oriented polarization position angle 
resulting from scattering prevent the $\psi$ value from stabilizing, resulting in a chaotic signal. This behavior is strengthened due to the edge-on 
orientation of the exoplanet's orbit. As the illuminated side of HD~189733b, producing the scattering-induced polarization signal, is facing the 
star, an observer situated on Earth no longer detects the bright side of the atmosphere and the signal is almost canceled. Only photons that 
have scattered on the top of the atmosphere carry a non-random polarization signal. Since most of the radiation comes from HD~189733A, 
the resulting polarization signal is highly diluted and thus $\psi$ randomly oscillates. 

Another variation of $\psi$ is detected when the planet passes behind HD~189733A: at the ingress (egress) point, the polarization position angle 
increases (decreases) by $\sim$~30$^{\circ}$, due to the partial obscuration of the reprocessing target. The inclination of the system and the partial 
obscuration of the planetary disk suppresses the averaging of polarization and results in $\psi$ variations before a pure cancellation of polarization 
due to the planet disappearance. The amplitude of the fluctuation of $\psi$ is directly correlated with the orbital inclination of the exoplanet and 
could be, in principle, used to estimate the impact parameter of the transit.

\section{Is direct detection of the exoplanet in X-rays possible?}
\label{Discussion}

We presented the first quantification of the amount of X-ray flux reprocessed by the Hot Jupiter exoplanet HD~189733b, along with the first 
estimation of the net polarization a future X-ray polarimeter could detect. 

\subsection{Scattered X-ray flux}

We find that the 0.7~keV flux of HD~189733b is three to five orders of magnitude lower than its primary main sequence K1.5~V star, depending on the 
orbital phase of the exoplanet. Considering a stellar flux of 2.9$\times$10$^{-13~}$erg~cm$^{-2}$~s$^{-1}$ at 0.7~keV and at a distance of 19.45~pc, it 
implies that the reprocessed flux from HD~189733b is lower than 1.0$\times$10$^{-16~}$erg~cm$^{-2}$~s$^{-1}$ at best before ingress and after egress. 

Since the planetary emission cannot be spatially resolved from the stellar emission in X-rays (angular separation lower than $1\farcs6$~mas), one 
can only look for flux modulation with orbital phase. As a guideline, we can use a threshold $f^\mathrm{limit}_\mathrm{p}$ of the planetary relative 
flux, $f_\mathrm{p}$, to divide the X-ray observation of $N$ planetary orbits (of period $P$ in seconds) in two parts. During the orbital phase 
$\Delta\phi$, where $f_\mathrm{p}$ is lower than $f^\mathrm{limit}_\mathrm{p}$, the X-ray count rate of HD~189733 is 
$CR_\mathrm{low}=<\!CR\!>(1+<\!f_\mathrm{p}\!>)$ with $<\!CR\!>$ the count rate of the host star and $<\!f_\mathrm{p}\!>$ the average planetary 
relative flux on $\Delta\phi$. During the orbital phase $1-\Delta\phi$, where $f_\mathrm{p}$ is larger than $f^\mathrm{limit}_\mathrm{p}$, the 
X-ray count rate of HD~189733 is $CR_\mathrm{high}=<\!CR\!>(1+\alpha <\!f_\mathrm{p}\!>)$, with $\alpha$ the average increase of the relative 
planetary flux. The increase of count rate $\Delta CR \equiv CR_\mathrm{high}-CR_\mathrm{low}$ is statistically significant if $\Delta CR \ge n \sigma$ 
with $n=4$ and assuming Poisson noise $\sigma^2=(CR_\mathrm{low}/\Delta\phi+CR_\mathrm{high}/(1-\Delta\phi))/(NP)$. Straightforward algebra leads to: 

\begin{equation}
N \gtrsim n^2/(f^2(\alpha-1)^2\Delta\phi(1-\Delta\phi)P<\!CR\!>).
\end{equation}

Using $<\!CR\!>=0.188$~pn~count~s$^{-1}$ (Sect.~\ref{Model:Corona}), we find a minimum value of $13,315$ orbits for $f^\mathrm{limit}_\mathrm{p}=7.5\times10^{-5}$, 
$\alpha=8.7$, and $\Delta\phi=0.89$. Since the needed accuracy requires a total exposure time of 81~years, the modulation of the X-ray flux with 
the orbital phase cannot be observed.

\subsection{X-ray polarization}

Fig.~\ref{Fig:Results} shows that the continuum polarization degree resulting from the interaction of stellar photons with the inner layers of 
the HD~189733b's atmosphere is very low, ranging from 0 to 0.003~\%. Due to strong dilution by the unpolarized primary source, this level 
of polarization corresponds to an amplitude of $\sim$~10$^{-4}$. Coupled to a low X-ray luminosity, such degree of polarization is, by any standard, 
by far too low to be detected with the current generation of X-ray polarimeters. Such statement includes the Gas Pixel Detector technology \citep{Soffitta2013}, 
Time Projection Chamber polarimeter concepts \citep{Black2010}, and polarimeters based on Bragg diffraction and Thomson scattering \citep{Kaaret1989}. 
Note that most of those polarimeter concepts are also limited to energies higher than 0.7~keV.

\section{Conclusions}
\label{Conclusion}

In this paper, we adopted the \cite{Salz2016}'s model of the atmosphere of HD 189733b to run a Monte-Carlo radiative transfer code to 
estimate the amount of X-ray photons emitted from the corona of HD~189733A reprocessed on this Hot-Jupiter like exoplanet. We obtained the 
predictive X-ray transit light curve of HD~189733A and quantified the flux and polarization emerging from scattering onto the atmospheric 
layers.

We have found that the maximum depth of the planetary transit on the geometrically thick and optically thin corona of HD~189733A
in the 0.25--2~keV energy band is $\sim$1.7\%. With the assumption that adding metals in the atmosphere would not change dramatically 
the density-temperature profile, we have estimated that the maximum transit depth is little sensitive to the metal abundance. Since 
the maximum depth of the X-ray transit of HD~189733b estimated from the current modeling of its atmosphere is small, forthcoming deeper
observations with the current X-ray observatories may not have the required X-ray sensitivity to accurately constrain the transit shape. 
The next generation of large X-ray mission from the ESA, Athena, thanks to its larger mirror area (e.g., the proposed Athena+ mission had 
at 1~keV a mirror area about 14 times larger than the FM2-pn mirror of XMM-Newton; \citealt{Nandra2013,Jansen2001}) should allow deeper 
investigations of the X-ray planetary transits of HD~189733A \citep{Branduardi2013}.

We have quantified the amount of flux reprocessed onto the Hot Jupiter surface. The exoplanet's flux is three to five orders of 
magnitude fainter than its primary star, with maximums at egress and ingress points due to the asymmetrical scattering phase function 
of Compton and Rayleigh scattering. At most, the reprocessed flux is lower than 1.0$\times$10$^{-16~}$erg~cm$^{-2}$~s$^{-1}$. 
A detection of the flux modulation with orbital phase is in pratical not achievable since requiring tremendous exposure time.

The reprocessed flux is associated with a very weak, diluted, amount of linear polarization. The polarization degree being less than 
to 0.003~\%, it is impossible to measure it with the current technology of X-ray polarimeters. 

However, the direct detection of the X-rays scattered by HD~189733b might be considered in the future with the possible advent of 
interferometric facilities in X-rays, e.g., \emph{the Black Hole Mapper} visionary-mission with (sub)microarcsecond resolution (see 
\emph{NASA Astrophysics in the Next Three Decades}, \citealt{Kouveliotou2014}).

\acknowledgements
This research has been supported by the Academy of Sciences of the Czech Republic, the COST Action MP1104 ``Polarization as a tool 
to study the Solar System and beyond'', and the European Union Seventh Framework Programme (FP7/2007-2013) under grant agreement 
no.\ 312789 ``StrongGravity''. FM is grateful to Sylvain Bugier (\textcolor{cyan}{sylvain.bugier@gmail.com}) for his artwork 
of the HD~189733 system.


\end{document}